\documentclass[final,3p]{elsarticle}

\usepackage{graphicx}
\usepackage{color}
\usepackage{epsfig}

\usepackage{amssymb}
\usepackage{amsmath}
\usepackage{amsthm}
\usepackage{booktabs}
\usepackage{latexsym}
\usepackage{eucal}
\usepackage{multirow}
\usepackage{upref}
\usepackage{hyperref}
\usepackage{subfigure}
\usepackage{ulem}
\usepackage{lettrine}
\usepackage{verbatim}
\usepackage{array}
\usepackage{varwidth}

\usepackage{framed} 
\usepackage{multicol} 

\usepackage[compatible]{nomencl} 
\makenomenclature

\renewcommand*\nompreamble{\begin{multicols}{2}}
\renewcommand*\nompostamble{\end{multicols}}

\biboptions{square,round}

\journal{International Journal of Hydrogen Energy}

\begin{document}

\begin{frontmatter}

\title{Computing supersonic non-premixed turbulent combustion by an SMLD flamelet progress variable model}

\author[dimeg,cemec]{A. Coclite}
\ead{alessandro.coclite@poliba.it}

\author[cira,cemec]{L. Cutrone}
\ead{l.cutrone@cira.it}

\author[lfa]{M. Gurtner}
\ead{gurtner@lfa.mw.tum.de}

\author[dimeg,cemec]{P. De Palma}
\ead{pietro.depalma@poliba.it}

\author[lfa]{O. J. Haidn}
\ead{haidn@lfa.mw.tum.de}

\author[dimeg,cemec]{G. Pascazio\corref{cor}}
\ead{giuseppe.pascazio@poliba.it}

\cortext[cor]{Corresponding author}

\address[dimeg]{Dipartimento di Meccanica, Matematica e Management (DMMM), Politecnico di Bari, Via Re David 200 -- 70125 Bari, Italy}

\address[cemec]{Centro di Eccellenza in Meccanica Computazionale (CEMeC), Politecnico di Bari, Via Re David 200 -- 70125 Bari, Italy}

\address[cira]{Centro Italiano Ricerche Aerospaziali (CIRA), via Maiorise -- 81043 Capua, Italy}

\address[lfa]{Institute for Flight Propulsion (LFA), Technische Universit\"{a}t M\"{u}nchen, Boltzmannstrasse 15 -- 85748, Garching, Germany}

\begin{abstract}

This paper describes the numerical simulation of the NASA Langley Research Center supersonic $H_{2}$-Air combustion chamber performed using two approaches to model the presumed probability density function (PDF) in the flamelet progress variable (FPV) framework. The first one is a standard FPV model, built presuming the functional shape of the PDFs of the mixture fraction, $Z$, and of the progress parameter, $\Lambda$. 
In order to enhance the prediction capabilities of such a model in high-speed reacting flows, a second approach is proposed 
employing the statistically most likely distribution (SMLD) techcnique to presume the joint PDF of 
$Z$ and $\Lambda$, without any assumption about their behaviour. 
The standard and FPV-SMLD models have been developed using the low Mach number assumption. In both cases,
the temperature is evaluated by solving the total-energy conservation equation, providing a more suitable approach for 
the simulation of supersonic combustion. 
By comparison with experimental data, the proposed SMLD model is shown to provide a clear improvement with respect to the standard FPV model, especially in the auto-ignition and stabilization regions of the flame.

\end{abstract}


\end{frontmatter}

\begin{thenomenclature}
\nomgroup{A}
\item [{$C$}]\begingroup Progress variable\nomeqref {0}\nompageref{1}
\item [{$D_\phi$}] \begingroup Diffusivity of $\phi$ \nomeqref {0}\nompageref{1}
\item [{$H$}] \begingroup Total enthalpy \nomeqref {0}\nompageref{1}
\item [{$k$}] \begingroup Turbulent kinetic energy \nomeqref {0}\nompageref{1}
\item [{$Ma$}] \begingroup Mach number of the flow \nomeqref {0}\nompageref{1}
\item [{$p$}] \begingroup Pressure\nomeqref {0}\nompageref{1}
\item [{$\widetilde{P}(x)$}] \begingroup Density-weighted probability density function \nomeqref {0}\nompageref{1}
\item [{$\widetilde{P}_{SML,2}(x)$}] \begingroup Density-weighted probability density function evaluated as the statistical most likely distribution with the second order moments \nomeqref {0}\nompageref{1}
\item[{$q$}]\begingroup Heat flux \nomeqref {0}\nompageref{1}
\item[{$\dot{q}_{react}$}]\begingroup Heat release rate \nomeqref {0}\nompageref{1}
\item [{$R$}] \begingroup Gas constant \nomeqref {0}\nompageref{1}
\item [{$Re$}] \begingroup Reynolds number of the flow \nomeqref {0}\nompageref{1}
\item [{$T$}] \begingroup Temperature  \nomeqref {0}\nompageref{1}
\item [{$Tu$}] \begingroup Turbulence intensity \nomeqref {0}\nompageref{1}
\item [{$u$}] \begingroup Velocity \nomeqref {0}\nompageref{1}
\item [{$Y_\phi$}] \begingroup Mass fraction of $\phi$ \nomeqref {0}\nompageref{1}
\item [{$Z$}]\begingroup Mixture fraction\nomeqref {0}\nompageref{1}

\item [{$\beta(x)$}] \begingroup $\beta$-distribution \nomeqref {0}\nompageref{1}
\item [{$\Gamma$}]\begingroup Euler function \nomeqref {0}\nompageref{1}
\item [{$\delta(x)$}] \begingroup Dirac distribution \nomeqref {0}\nompageref{1}
\item [{$\Lambda$}]\begingroup Progress parameter\nomeqref {0}\nompageref{1}
\item [{$\mu$}] \begingroup Dynamic viscosity \nomeqref {0}\nompageref{1}
\item [{$\mu_x$}] \begingroup Lagrangian multiplier \nomeqref {0}\nompageref{1}
\item [{$\nu$}] \begingroup Kinematic viscosity \nomeqref {0}\nompageref{1}

\item [{$\rho$}]\begingroup Density \nomeqref {0}\nompageref{1}
\item [{$\phi$}]\begingroup Generic thermo-chemical quantity\nomeqref {0}\nompageref{1}
\item [{$\widetilde{\phi}$}]\begingroup Favre-averaged value of $\phi$\nomeqref {0}\nompageref{1}
\item [{$\phi ''$}]\begingroup Fluctuation of $\phi$ in the Favre-averaging process\nomeqref {0}\nompageref{1}
\item [{$\widetilde{\phi''^2}$}]\begingroup Variance of $\phi$\nomeqref {0}\nompageref{1}
\item [{$\overline{\phi}$}]\begingroup Reynolds-averaged value of $\phi$\nomeqref {0}\nompageref{1}
\item [{$\phi '$}]\begingroup Fluctuation of $\phi$ in the Reynolds-averaging process\nomeqref {0}\nompageref{1}
\item[{$\Phi$}] \begingroup Error function\nomeqref {0}\nompageref{1}
\item [{$\chi$}] \begingroup Scalar dissipation rate \nomeqref {0}\nompageref{1}
\item [{$\chi_{st}$}] \begingroup Value of $\chi$ at the stoichiometric point \nomeqref {0}\nompageref{1}
\item [{$\omega$}] \begingroup Turbulent kinetic energy specific dissipation rate \nomeqref {0}\nompageref{1}
\item [{$\dot{\omega}_\phi$}] \begingroup Production therm of $\phi$ \nomeqref {0}\nompageref{1}


\end{thenomenclature}

\section{Introduction}
\label{intro}

In the last years, the development of propulsion systems based on air-breathing engines for
super-/hyper-sonic flying vehicles has fostered the study of supersonic combustion, see, e.g., the review of Cecere et al.~\citep{Cecere2014}.
In these systems,
hydrogen is one of the preferred fuel because of its properties in terms of very short ignition delay time and high energy per unit weight. 
The investigation of hydrogen supersonic combustion presents significant difficulties and high costs either following the experimental approach or the numerical one.
In fact, in supersonic combustion, the mixing time scales are comparable to  $H_{2}$--Air reaction time scales~\cite{cheng}. 
Moreover, high-Reynolds-number turbulent combustion is a formidable multi-scale problem, where the interaction between chemical kinetics, molecular, and turbulent transport occurs over a wide range of length and time scales. These features pose severe difficulties in the analysis and comprehension of the basic phenomena involved in supersonic combustion.\\ 
Concerning the numerical approach, in recent years, the need for efficient tools having affordable computational costs has driven the research towards: i) studying turbulent combustion in order to understand the interaction between turbulence and chemistry~\cite{heinz2010,fan2012,jin2013dns,Cecere2012}; ii) improving kinetic schemes to describe the combustion process~\cite{boivin2012supersonic,williams2008,Bezgin2013}.
Higher accuracy can be achieved by employing models based on detailed kinetic mechanisms, 
but this usually leads to prohibitively expensive calculations. Therefore, 
reduced models are often employed to condensate the reaction mechanisms and cut down the computational costs~\cite{boivin2012supersonic}. 
Simplified approaches to combustion modelling
have been proposed to further reduce the number of equations to be solved; for instance,
the reduction of the chemical scheme in intrinsic low dimensional manifolds~(ILDM)~\cite{maas}; the flamelet-based approaches such as the flamelet-progress variable~(FPV)~\cite{pierce} or flame prolongation of ILDM~(FPI)~\cite{laminarhydrogen}; and flamelet generated manifolds approach~(FGM)~\cite{oijen}. 

The present work is based on the FPV model for non-premixed flames. Standard steady FPV models are built under the low Mach number 
hypothesis~\cite{pitsch98,carbonell2009,knudsen2009,knudsen2012}, computing the combustion process at a uniform given pressure. 
Obviously, in supersonic combustion, density variations due to the dynamics of the flow cannot be neglected. 
For this reason, here we employ the modified FPV approach proposed by Oevermann~\cite{Oever}, and recently employed by other authors~\cite{hou2014}, where the fluid temperature is not obtained from the flamelet libraries as in the standard model, but is calculated by solving the full set of conservation equations for compressible flows, including the energy equation. 
More recent works have proposed an extension of FPV combustion models to compressible flows by using a source term linearly rescaled 
with the pressure value and a perturbation of the low-Mach number flamelet accounting for compressibility~\cite{terrapon2012,pitsch2014}.

For the case of non-premixed combustion of interest here, mixing must bring reactants into the reaction zone so as to activate and maintain the combustion process. 
Such flames are characterized by a local balance between diffusion and reaction~\cite{peters}.
The basic assumption of the flamelet model is that each element of the flame front can be described as a small laminar flame, also called {\it flamelet}.
Therefore, for a steady flow, the flame structure can be described as a function of the mixture fraction, $Z$, and of the progress parameter, $\Lambda$~\cite{piercemoin2004}. 
For turbulent combustion, a probability density function (PDF) is needed to compute the mean value and the variance of the thermo-chemical variables.  
The definition of such a PDF is critical since it has a strong impact on the solution.
The aim of this work is to study the applicability of the statistically most likely distribution (SMLD)~\cite{pope} approach to model joint-PDF of $Z$ and $\Lambda$ in the case of supersonic combustion. The proposed joint-SMLD approach is very interesting since it represents a good compromise between computational costs and accuracy level.
The results obtained using the proposed model are validated versus experimental data and compared with numerical results available in the literature as well as with those obtained using the standard FPV model.

This work is organised as follow:  sections~\ref{fpv} and~\ref{goveq} provide the theoretical description and some details of the
numerical discretization for the two models. 
The comparison between their numerical results for the simulation of the NASA Langley Research Center supersonic hydrogen flame are presented in section~\ref{results}
along with reference numerical and experimental data. Finally, some conclusions are provided.

\section{The flamelet progress variable models}
\label{fpv}
For the case of non-premixed combustion of interest here, the basic assumptions of the flamelet model are fulfilled
for sufficiently large Damk\"{o}hler number, $Da$. In fact, when the reaction zone thickness is very thin with respect to the Kolmogorov
length scale, turbulent structures are unable to penetrate into the reaction zone and cannot destroy the laminar flame structure. 
Effects of turbulence only result in a deformation and
straining of the flame sheet and locally the flame structure can be described as function of the mixture fraction,  $Z$,
the scalar dissipation rate, $\chi$, and the time. The scalar dissipation rate, $\chi=2 D_Z(\nabla Z)^2$,
is a measure of the gradient of the mixture fraction representing the molecular diffusion of the species in the flame,
$D_Z$ being the molecular diffusion coefficient of the chemical species.
Therefore, the entire flame behaviour can be obtained as a combination of solutions of the laminar flamelet equation.
In the present work we consider a further simplification assuming a steady flamelet behaviour, so that chemical effects are entirely
determined by the value of $Z$, whereas $\chi$ describes the effects of the flow on the flame structure according to
the following steady laminar flamelet equation (SLFE) for the generic variable $\phi$:
\begin{equation}
\label{slfe}
{-\rho\frac{\chi}{2}\frac{\partial^2 \phi}{\partial Z^2}=\dot\omega_\phi}.
\end{equation}
In equation \eqref{slfe},  $\rho$ is the density and  $\dot\omega_\phi$ is the source term related to $\phi$~\cite{piercemoin2004}, different from zero in the case of finite rate chemistry. 
In particular, in this work, the FPV model proposed by Pierce and Moin~\cite{pierce,piercemoin2004} is employed to evaluate all of the thermo-chemical quantities involved in the combustion process. 
This approach is based on the parametrization of the generic thermo-chemical quantity, $\phi$, in terms of the mixture fraction, $Z$, and of the progress parameter, $\Lambda$, instead of $\chi$:
\begin{equation}
\label{phi}
{\phi=F_\phi(Z,\Lambda)}\, .
\end{equation}
Using such a parameter, independent of the mixture fraction, one can uniquely identify each flame state along the stable and unstable branches of the S-shaped curve. A suitable definition of $\Lambda$ leads to a dramatic simplification of the presumed PDF closure model. On the other hand, the solution of the transport equation for $\Lambda$ is quite complex since it requires non-trivial modelling of several unclosed terms~\cite{ihmea}. In order to overcome such a difficulty, the progress parameter is derived from a reaction progress variable, $C$, such as the temperature or a linear combination of the main reaction products, whose behaviour is governed by a simpler transport equation. Therefore, a transport equation for $C$ is solved and the flamelet library is parametrized in terms of $Z$ and $C$. Requiring that the transformation between $\Lambda$ and $C$ be bijective, from equation~\eqref{slfe} one has
\begin{equation}
\Lambda=F^{-1}_C(Z,C)\, ,
\end{equation} 
and any thermo-chemical variable can be expressed as:
\begin{equation}
\phi=F_\phi(Z,F^{-1}_C(Z,C))\, .
\end{equation}
The choice of the progress variable is not unique and some recent works discuss in details this issue proposing a procedure for its optimal 
selection~\cite{ihmejcp2012,cuenotCF2012,vervischCF2013}. A suitable definition for the progress variable 
is the sum of the mass fraction of the main products~\cite{ihmejcp2012}; for hydrogen combustion~\cite{pitsch2014}:
\begin{equation}
{C=Y_{H_2O}}.
\end{equation} 
A stretching with respect to the minimum and maximum conditioned value of $C$ over the mixture fraction is made, leading to the following form of the progress parameter:
\begin{equation}
\label{normalization}
\Lambda = \frac{{C} - {C_{Min}|Z}}{{C_{Max}|Z} - {C_{Min}|Z}}\, .
\end{equation}
Equation~(\ref{phi}) is taken as the solution of the SLFE~\eqref{slfe}. 
Even if cases in which there is not a unique mapping of this solution in function of $Z$ and $\Lambda$ are reported in the literature~\cite{pierce, piercemoin2004}, such events are excluded from the solution family composing the flamelet library, since they are very close to the equilibrium limit.
Since flamelet libraries are computed in advance and are assumed to be independent of the flow field, one has to model
the dependence of $\chi$ on $Z$~\cite{pitsch98,pitschchenpeters98,kimwilliams93}. 
In this work, the functional form of $\chi(Z)$ has been taken from an idealized flow configuration, as proposed by
Peters~\cite{peters84}; the distribution of the scalar dissipation rate in a counterflow diffusion flame, $\Phi(Z)$,  is employed, scaled in the following way:
\begin{equation}
\label{chi_st}
{\chi(Z)=\chi_{st}\frac{\Phi(Z)}{\Phi(Z_{st})}},
\end{equation}
where $\chi_{st}$ and $Z_{st}$ are evaluated at the stoichiometric point~\cite{peters84}.

From equation \eqref{phi} one can derive the generic thermo-chemical quantity $\phi$.  When a turbulence model is used, the Favre-averaged
value of  $\phi$ and its variance are given as:
\begin{equation}
\label{media}
{\widetilde\phi=\int\int F_\phi(Z,\Lambda)\widetilde{P}(Z,\Lambda)dZ d\Lambda},
\end{equation}
\begin{equation}
\label{varianza}
{\widetilde{\phi''^2}=\int\int (F_\phi(Z,\Lambda)-\widetilde\phi)^2\widetilde{P}(Z,\Lambda)dZd\Lambda}.
\end{equation}
In the above equations,  $\widetilde P(Z,\Lambda)$ is the density-weighted PDF,
\begin{equation}
{\widetilde P(Z,\Lambda)=\frac{\rho P(Z,\Lambda)}{\overline{\rho}}},
\end{equation}
$P(Z,\Lambda)$ is the joint PDF and $\overline{\rho}$ is the Reynolds-averaged density. As usual, $\phi$ can be decomposed as:
\begin{equation}
\phi=\widetilde \phi+ \phi'',
\qquad
\widetilde \phi = \frac{\overline{\rho \phi }}{\overline{\rho}}
\end{equation}
and
\begin{eqnarray}
\rho&=&\overline{\rho} +\rho',
\end{eqnarray} 
where $\phi''$ and $\rho'$ are the fluctuations.
The joint-PDF, $\widetilde{P}(Z,\Lambda)$, plays a crucial role in the definition of the model, affecting both its accuracy and computational costs.
Moreover, the choice of such a function is not straightforward because of the unknown statistical behaviour of the two variables $Z$ and $\Lambda$~\cite{ihmea} and its definition is still an open problem whose solution is being pursued by several 
researches~\cite{ihmeal2005,ihmeb,DeMeesterCF2012,AbrahamPoF2012}.
The aim of this work is to validate a more general model based on the statistically most likely distribution (SMLD)~\cite{pope} for the joint PDF of $Z$ and 
$\Lambda$~\cite{cocliteFTC2014}. 
The performance of such a combustion model are assessed by computing a hydrogen-air supersonic flame and comparing the
results with those obtained using the standard FPV model.
In the present work, the Reynolds-Averaged Navier--Stokes equations with $k$-$\omega$ turbulence model~\cite{luigi} are solved and
both the standard-FPV and the FPV-SMLD models  employ the total energy conservation equation to evaluate the temperature field in order to
improve the simulation of compressible reacting flows.

\subsection{Presumed probability density function model}

In this section, the standard FPV model~\cite{piercemoin2004} (called here model A) and the FPV-SMLD model (called here model~B)~\cite{cocliteFTC2014} are briefly described. 

The evaluation of the average quantities in equations \eqref{media} and \eqref{varianza} requires the PDF to be known or somehow presumed.
Such a PDF establishes the statistical correlation between $Z$ and $\Lambda$. 
Employing the Bayes' theorem, 
\begin{equation}
\label{bayes}
{\widetilde{P}(Z,\Lambda)=\widetilde P(Z)\widetilde P(\Lambda|Z)} \, ,
\end{equation}
one usually presumes the functional shape of the marginal PDF of $Z$ and of the conditional PDF of $\Lambda|Z$.
In model~A, the basic assumption is the statistical independence between $Z$ and $\Lambda$, so that $\widetilde{P}(Z,\Lambda)=\widetilde P(Z)\widetilde P(\Lambda)$.
Furthermore, the statistical behaviour of the mixture fraction is described by a $\beta$-distribution.
In fact, even though the definition of $\widetilde P(Z)$ is still an open question~\cite{pope}, 
it has been shown by several authors that the mixture fraction behaves like a passive scalar whose statistical distribution can be approximated by a $\beta$~function~\cite{cook,jimenez,wall}. 
The two-parameter family of the $\beta$-distribution in the interval $x\in [0,1]$ is given by:
\begin{equation}
\label{beta}
{\beta(x;\widetilde{x},\widetilde{x''^2})=x^{a-1}(1-x)^{b-1}\frac{\Gamma(a+b)}{\Gamma(a)\Gamma(b)}},
\end{equation}
where $\Gamma(x)$ is the Euler function and $a$ and $b$ are two parameters related to $\widetilde x$ and $\widetilde{x''^2}$
\begin{equation}
\label{aeb}
{a=\frac{\widetilde x(\widetilde x- \widetilde{x}^2-\widetilde{x''^2})}{\widetilde{x''^2}}, \ \ \ b=\frac{(1-\widetilde x)(\widetilde x-\widetilde{x}^2-\widetilde{x''^2})}{\widetilde{x''^2}}}.
\end{equation} 
Moreover, $\widetilde P(\Lambda)$ is chosen as a Dirac distribution, implying a great simplification in the theoretical framework.
With these assumptions, the Favre-average of a generic thermo-chemical quantity is given by:
\begin{equation}
\label{phidelta}
{\widetilde\phi=\int\int F_\phi(Z,\Lambda)\widetilde \beta(Z)\delta(\Lambda-\widetilde{\Lambda})dZdC=\int F_\phi(Z,\widetilde \Lambda)\widetilde \beta(Z)dZ}.
\end{equation} 
Therefore, in addition to the conservation and turbulence model equations, one has to solve only three transport equations (for $\widetilde Z$, $\widetilde{Z''^2}$ and $\widetilde C$) to evaluate all of the thermo-chemical quantities, thus avoiding the expensive solution of one transport equation for each chemical species. 

Model~B, based on the SMLD approach to model the joint PDF, does not need any assumption about the form of $\widetilde{P}(Z,\Lambda)$.
Following such an approach, the probability distribution can be evaluated as a function of an arbitrary number of moments of $Z$ and $\Lambda$.
It is noteworthy that, even though equation~\eqref{slfe} is based on the assumption that $Z$ and $\Lambda$ are independent, one can properly take into account the statistical correlation between $Z$ and $\Lambda$ employing the SMLD joint-PDF in the evaluation of the effects of turbulence~\cite{ihmeal2005}. 

In this work, the first two moments of the joint probability density function $\widetilde P(\vec x)$, where $\vec x=(Z,\Lambda)^T$, are assumed to be known; 
therefore, the joint-PDF reads~\cite{cocliteFTC2014} :
\begin{multline}
\label{smld2}
\widetilde P_{SML,2}(Z,\Lambda)= \frac{1}{\mu_0}\exp\Bigl\{-\Bigl[\mu_{1,1} (Z-\widetilde Z)+\mu_{1,2}(\Lambda-\widetilde \Lambda)\Bigr]\\-\frac{1}{2}\Bigl[\mu_{2,11}(Z-\widetilde Z)^2+\mu_{2,12}(Z-\widetilde Z)(\Lambda-\widetilde \Lambda)+\mu_{2,21}(\Lambda-\widetilde \Lambda)(Z-\widetilde Z)+\mu_{2,22}(\Lambda-\widetilde \Lambda)^2 \Bigl]
\Bigr\}.
\end{multline}
In the equation above, $\mu_0$ is a scalar, $\vec{\mu_1}$ is a two~-~component vector,
and $\overleftrightarrow{\mu_2}$ is a square matrix of rank two:
\begin{eqnarray}
\label{moltiplicatori}
\mu_0&=&\int d\vec x \widetilde P_{SML,2}(\vec x) ,\\
\label{moltiplicatori1}
-\mu_{1,i}&=&\int d\vec x\partial_{x_i} \widetilde P_{SML,2}(\vec x)=\beta(1;\widetilde\xi_i,\widetilde{\xi_i''^2})-\beta(0;\widetilde \xi_i,\widetilde{\xi_i''^2}) ,\\
\label{moltiplicatori2}
\delta_{kl}-\mu_{2,kn}\ \widetilde{\xi'_n\xi'_l}&=&\int d\vec x \partial_{x_k}((x_l-\widetilde\xi_l)\widetilde P_{SML,2}(\vec x))=\beta(1;\widetilde \xi_k,\widetilde{\xi'_k\xi'_l})-\widetilde\xi_k\mu_{1,l} ,
\end{eqnarray} 
where $i$, $k$, $n$, and $l$ indicate the vector components; 
$\widetilde\xi_i$, $\widetilde\xi'_i$ and $\widetilde{\xi^{''2}_i}$ are the mean ($\widetilde{x_i}$), the fluctuation ($x_i-\widetilde{x_i}$) and the variance ($\widetilde{x_i''^2}$) of the $i$-th component of $\vec x$, respectively; finally, $\beta$ indicates the beta distribution function.
In model B, one has to solve four additional transport equations (for $\widetilde Z$, $\widetilde{Z''^2}$, $\widetilde C$, and $\widetilde{C''^2}$) 
to evaluate all of the thermo-chemical quantities.

\section{Governing equations}
\label{goveq}

\subsection{Flow equations and numerical solution procedure}

The numerical method developed in~\cite{luigi} has been employed to solve the steady-state RANS equations with $k$-$\omega$ turbulence closure. 
For an axisymmetric multi-component reacting compressible flow the system of governing equations can be written as:
\begin{equation}
\label{floweq}
{\partial_t \vec Q+\partial_{x} (\vec E-\vec E_{\nu})+\partial_{y} (\vec F-\vec F_{\nu})= \vec S},
\end{equation} 
where $t$ is the time variable; $x$ and $y$ are the axial and the radial coordinate, respectively;
$\vec Q$=($\overline \rho$,$\,\overline \rho \widetilde u_x$,$\,\overline \rho \widetilde u_y$,$\,\overline \rho \widetilde H-\widetilde p_t$,$\,\overline \rho  
\widetilde k$, $\,\overline \rho \widetilde \omega$,$\,\overline \rho \widetilde R_n$) is the 
vector of the conserved variables; $\vec E$, $\vec F$, and  $\vec E_v$, $\vec F_v$ are the inviscid and viscous flux vectors \cite{dsthesis}, respectively; $\vec S$ is the vector of the source terms; $\overline \rho$, $(\widetilde u_x,\widetilde u_y)$, $\widetilde H$ indicate the Reynolds-averaged value of density, the Favre-averaged values of velocity components and specific total enthalpy given by $\widetilde H =\widetilde h+\frac{1}{2}(\widetilde{u}_x^2+\widetilde{u}_y^2)+\frac{5}{3}\widetilde k$ with $\widetilde h$ accounting for the species enthalpy per unit mass, respectively; $\widetilde p_t = \widetilde p + \frac{2}{3}\widetilde k$,
$\widetilde p$ being the Favre-averaged value of pressure;  
$\widetilde k$ and $\widetilde \omega$ are the Favre-averaged values of the turbulence kinetic energy and of its specific dissipation rate;
$\widetilde R_n$ is a generic set of conserved variables related to the combustion model.
In this framework, $\widetilde R_n$ is the set of independent variables of the flamelet model, namely, $\widetilde Z$, ${\widetilde{Z''^2}}$,
$\widetilde C$, ${\widetilde{C''^2}}$ (see the following subsection).\\
The heat flux in the total energy equation, namely, 
\begin{equation}
\label{eq:HFlux}
q=-\rho c_p D_T\nabla T+\sum_{n=1}^{N_s} \rho V_nY_nh_n+\dot{q}_{react},
\end{equation}
is composed of three terms since the Dufour effect is neglected; $D_T$ is the thermal diffusivity, $c_p$ the specific heat at constant pressure;
the mass diffusion term is modelled by the Fick law considering $V_n=-D \frac{\nabla Y_n}{Y_n}$, assigning mixture diffusivity, $D$, to each species and so assuming unitary Lewis number~\cite{peters}.
The third term at the right hand side of the above equation represents the heat release rate:
\begin{equation}
\label{eq:HHR}
\dot{q}_{react}=\sum_{k=1}^N \Delta h^o_{f,k}\dot{\omega}_k\, ,
\end{equation}
where $k$ is the species index, $\Delta h^o_{f,k}$ is the mass formation enthalpy and $\dot{\omega}_k$ is the production rate of species $k$.

\subsection{Turbulent FPV transport equations}

For the case of turbulent flames, the solution of the SLFE, namely equation~\eqref{phi}, is expressed in terms of the Favre averages of $Z$ and $C$ and of their variance. Using model~A, one can tabulate all chemical quantities in terms of $\widetilde Z$, $\widetilde{Z^{''2}}$ and $\widetilde C$, since, due to the properties of the $\delta$-distribution, the model is independent of $\widetilde{C^{''2}}$. On the other hand, model~B expresses $\phi$ also in terms of $\widetilde{C^{''2}}$ 
and therefore an additional transport equation needs to be solved. 
In this case, the transport equations for the combustion model (included in equation~(\ref{floweq})) are written as:
\begin{eqnarray}
\label{zmean}
 \partial_t(\overline{\rho}\widetilde{Z})+\vec\nabla\cdot(\overline{\rho}\widetilde{\vec u}\widetilde{Z})&=&
\vec\nabla\cdot\Bigl[\bigl( D+
D_{\widetilde{Z}}^t\bigr)\overline{\rho}
\vec\nabla\widetilde{Z}\Bigr],\\
\label{zvar}
\partial_t(\overline{\rho}\widetilde{Z''^2})+\vec\nabla\cdot(\overline{\rho}\widetilde{\vec u}\widetilde{Z''^2})&=&
\vec\nabla\cdot\Bigl[\bigl( D+D_{\widetilde{Z''^2}}^t\bigr)\overline{\rho}\vec\nabla\widetilde{Z''^2}\Bigr]-\overline{\rho}\widetilde{\chi}+2\overline{\rho}D_{\widetilde Z}^t(\vec\nabla\widetilde{Z})^2,\\
\label{cmean}
\partial_t(\overline{\rho}\widetilde{C})+\vec\nabla\cdot(\overline{\rho}\widetilde{\vec u}\widetilde{C})&=&
\vec\nabla\cdot\Bigl[\bigl( D+D_{\widetilde{C}}^t\bigr)\overline{\rho}\vec\nabla\widetilde{C}\Bigr]+\overline{\rho}\overline{\dot\omega_C},\\
\label{cvar}
\partial_t(\overline{\rho}\widetilde{C''^2})+\vec\nabla\cdot(\overline{\rho}\widetilde{\vec u}\widetilde{C''^2})&=&
\vec\nabla\cdot\Bigl[\bigl( D+D_{\widetilde{C''^2}}^t\bigr)\overline{\rho}\vec\nabla\widetilde{C''^2}\Bigr]-\overline{\rho}\widetilde{\chi}_C+2\overline{\rho}D_{\widetilde C}^t(\vec\nabla\widetilde{C})^2+2\overline{\rho}\widetilde{C''\dot\omega''_C},
\end{eqnarray}
where $\widetilde{\chi}_C$ is modelled in terms of $\widetilde{Z''^2}$ and $\widetilde{C''^2}$~\cite{ihmeb}, namely $\widetilde{\chi}_C=\frac{\widetilde{Z''^2}\chi}{\widetilde{C''^2}}$, $D=\nu/Pr$ is the diffusion coefficient for all of the species; $\nu$ and $Pr$ are the kinematic viscosity and the Prandtl number, respectively; $D_{\widetilde Z}^t=D_{\widetilde{Z^{''2}}}^t=D_{\widetilde C}^t=D_{\widetilde{C''^2}}^t=\nu_t/Sc_t$ are the turbulent mass diffusion coefficients, ${Sc}_t$ being the turbulent Schmidt number equal to $0.8$; 
finally, $\dot{\omega}_{C}$ is the source term for the progress variable precomputed and tabulated in the flamelet library. 
At every iteration, the values of the flamelet variables are updated using equations~\eqref{zmean}-\eqref{cvar} and the Favre-averaged thermo-chemical quantities are computed, using equation \eqref{media}. Such solutions provide the mean mass fractions which are used to evaluate all of the transport properties of the fluid, namely the molecular viscosity, the thermal conductivity and the species diffusion coefficients.\\
The evaluation of the flamelet library has been performed using the detailed kinetic scheme proposed by Saxena and Williams~\cite{saxena2006}: $244$ sub-reactions upon $50$ species. The flamelet library has been computed over a grid with $250$ uniformly distributed points in the $\widetilde Z$ and $\widetilde C$ directions, and $50$ uniformly distributed points in the $\widetilde{Z''^2}$ and $\widetilde{C''^2}$ directions and a quadri-linear interpolation scheme is used.
The flamelet library has been evaluated considering a constant background pressure equal to $100\ kPa$ and boundary conditions for $Z$ and $C$ corresponding to the conditions of the air and $H_2$ streams. Indeed, the air stream is represented as $\widetilde Z=0$ and $\widetilde C=0$, while the fuel jet is given by $\widetilde Z=1$ and $\widetilde C=0$. 
The solution of the SLFE has been obtained by using the FlameMaster code~\cite{flamemaster}.

\subsection{Numerical scheme and boundary conditions}

A cell-centred finite volume space discretization is used on a multi-block structured mesh. The convective and viscous terms are discretized by the third-order-accurate Steger and Warming~\cite{steger} flux-vector-splitting scheme and by second-order-accurate central differences, respectively. An implicit time marching procedure is used with a factorization based on the diagonalization procedure of Pulliamm and Chaussee~\cite{pulliam}, employing a scalar alternating direction implicit (ADI) solution procedure~\cite{buelow97}. Steady flows are considered and the ADI scheme is iterated in the pseudo-time until a residual drop of at least five orders of magnitude for all of the conservation-law equations~\eqref{floweq} is achieved. Characteristic boundary conditions for the flow variables are imposed at inflow and outflow points.
In particular, a plug flow is imposed at the inlet points of the computational domain so as to match the experimental
conditions at the inlet section of the chamber, see table \ref{cheng_parameter}. Moreover,
$k$, $\omega$, and $\tilde R_n$ are assigned at inflow points, whereas they are linearly extrapolated at outflow points.
Finally, no slip and adiabatic conditions are imposed at walls, where
$k$ is set to zero and $\omega$ is evaluated as proposed in~\cite{menter}:
\begin{equation}
\label{eq:menter}
\omega = 60 \, \dfrac{\nu}{0.09 \, y_{n,1}^2 },
\end{equation}
where $y_{n,1}$ is the distance of the first cell center from the wall; the homogeneous Neumann
boundary condition is used for $\tilde R_n$ (non-catalytic wall).
Finally, symmetry conditions are imposed at the axis.
 
\section{Results and discussion}
\label{results}
\subsection{Description of the test case}

\begin{figure}
\begin{center}
\includegraphics[scale=0.3]{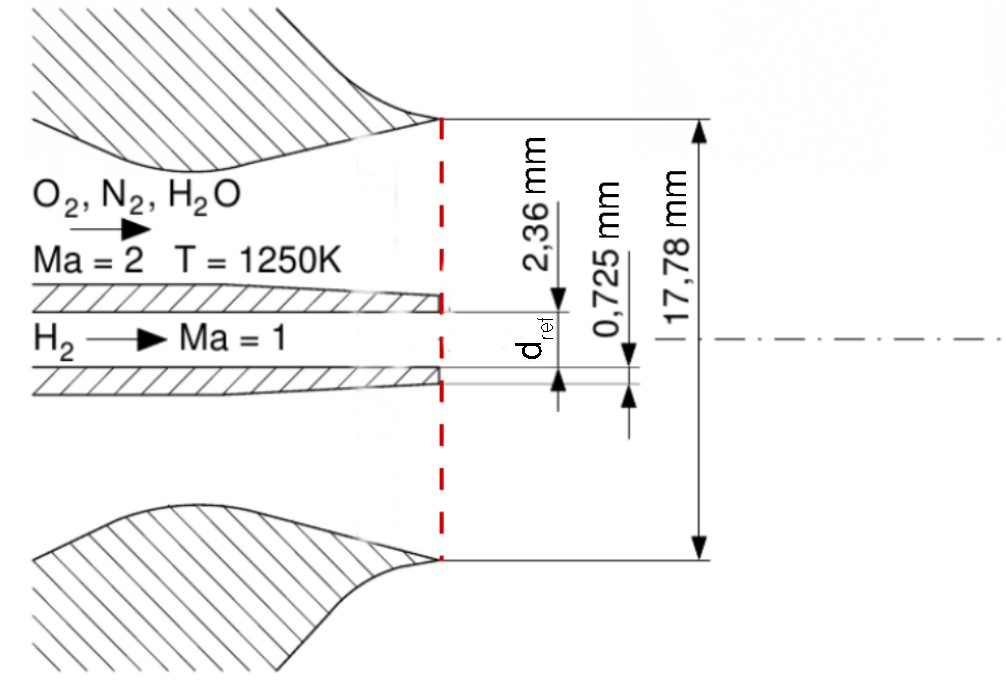}
\caption{Schematic of the Cheng's Flame burner. The red-dashed line represents the exit section of the burner and so the inlet section of the chamber.}
\label{cheng_geom}
\end{center}
\end{figure}

\begin{figure}
\begin{center}
\includegraphics[scale=0.2]{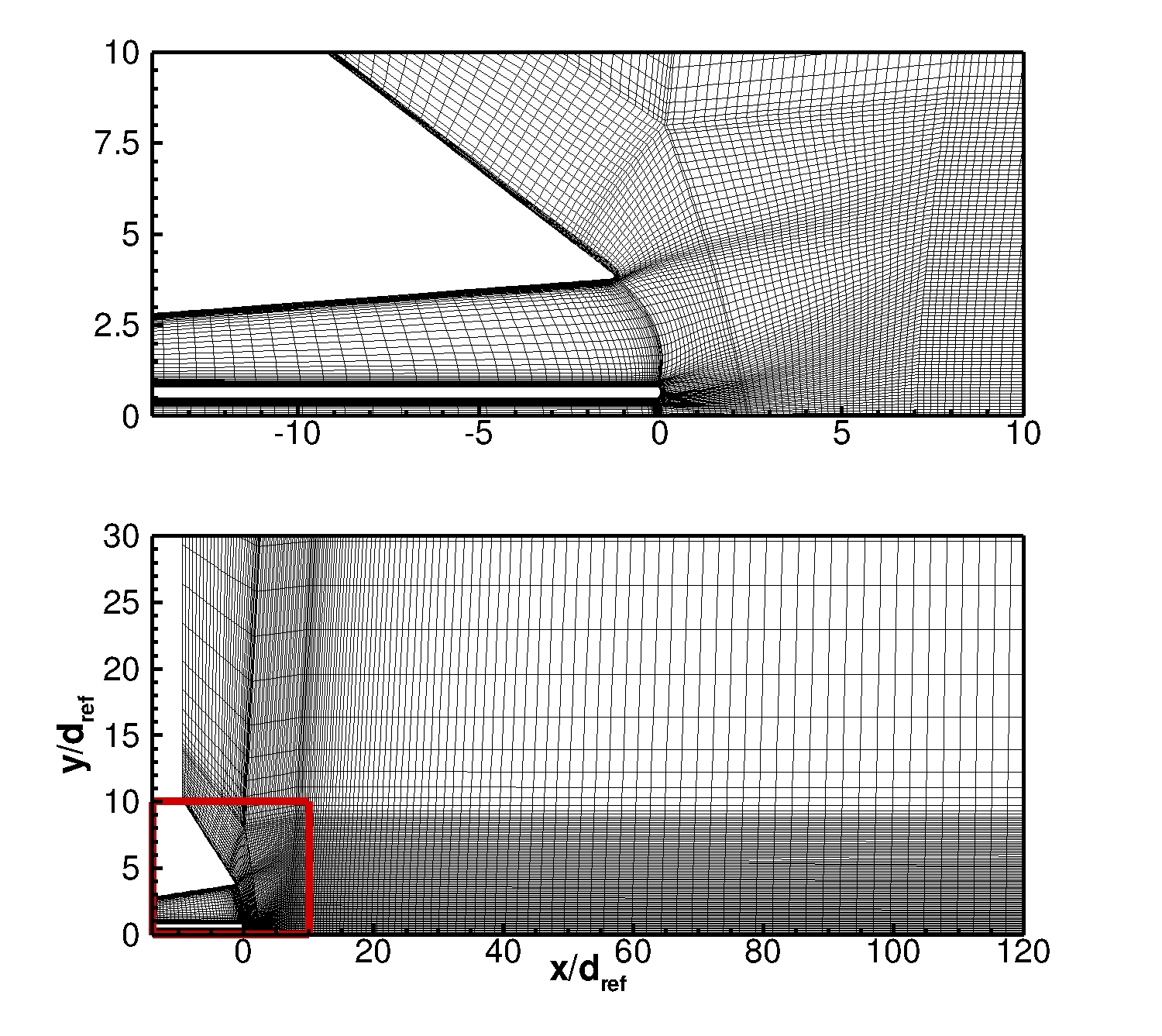}
\caption{Computational domain discretization (lower frame) and a detail of the injectors (upper frame).}
\label{cheng_grid}
\end{center}
\end{figure}

The hydrogen-air supersonic combustion burner studied by Cheng et al.~\cite{cheng1994} at the NASA Langley Research Center has been considered as a suitable test case for the proposed method.
At the inlet section of the chamber, the supersonic flow (see figure~\ref{cheng_geom}) is characterized by an annular axisymmetric jet of hot vitiated wet air at Mach number equal to $2$, average axial velocity of $1420\ m/s$, temperature of $1250\ K$ and pressure of $107\ kPa$. The values of the inner and outer diameters of the vitiated air annular duct are equal to $3.812\ mm$ and $17.78\ mm$, respectively. 
The composition of the vitiated air is: $Y_{O_2}=0.201$, $Y_{H_2O}=0.255$ and $Y_{N_2}=0.544$, produced by a pre-combustion at low temperature~\cite{cheng1994}.
The diameter of the fuel stream is $d_{ref}=2.362\ mm$, taken as the reference length.  
The hydrogen flow is estimated to be chocked with average axial velocity of $1780\ m/s$, temperature of $545\ K$, and pressure of $112\ kPa$. 
The operating conditions reported by Cheng et al.~\cite{cheng1994} are summarized in table~\ref{cheng_parameter} together with the turbulent intensity levels, $Tu_{in}$, and the turbulent viscosity, $\nu_{T_{in}}$.
The computational domain, shown in figure~\ref{cheng_grid}, is axisymmetric and includes the divergent part of the air nozzle, necessary to recover the correct flow quantities at the inlet section of the chamber;
it extends $150\ d_{ref}$ and $50\ d_{ref}$ along the axial and radial directions, respectively, 
and, after a grid refinement study, has been discretized using about $100000$ cells. 
The characteristic cell lengths at the exit of the divergent part of the nozzle are about $0.13\ mm$ and $0.05\ mm$ in the axial and radial directions, respectively.  
Numerical results obtained using the two models described in the previous sections are discussed and validated versus the experimental data of
Cheng et al.~\cite{cheng1994} and the numerical results of~\cite{boivin2012supersonic,boivin2014CST}. 
In particular, a detailed analysis of the flame structure is provided.
\subsection{Comparison between numerical and experimental data}
\begin{figure}
\begin{center}
\subfigure[Model~A]{\includegraphics[scale=0.15]{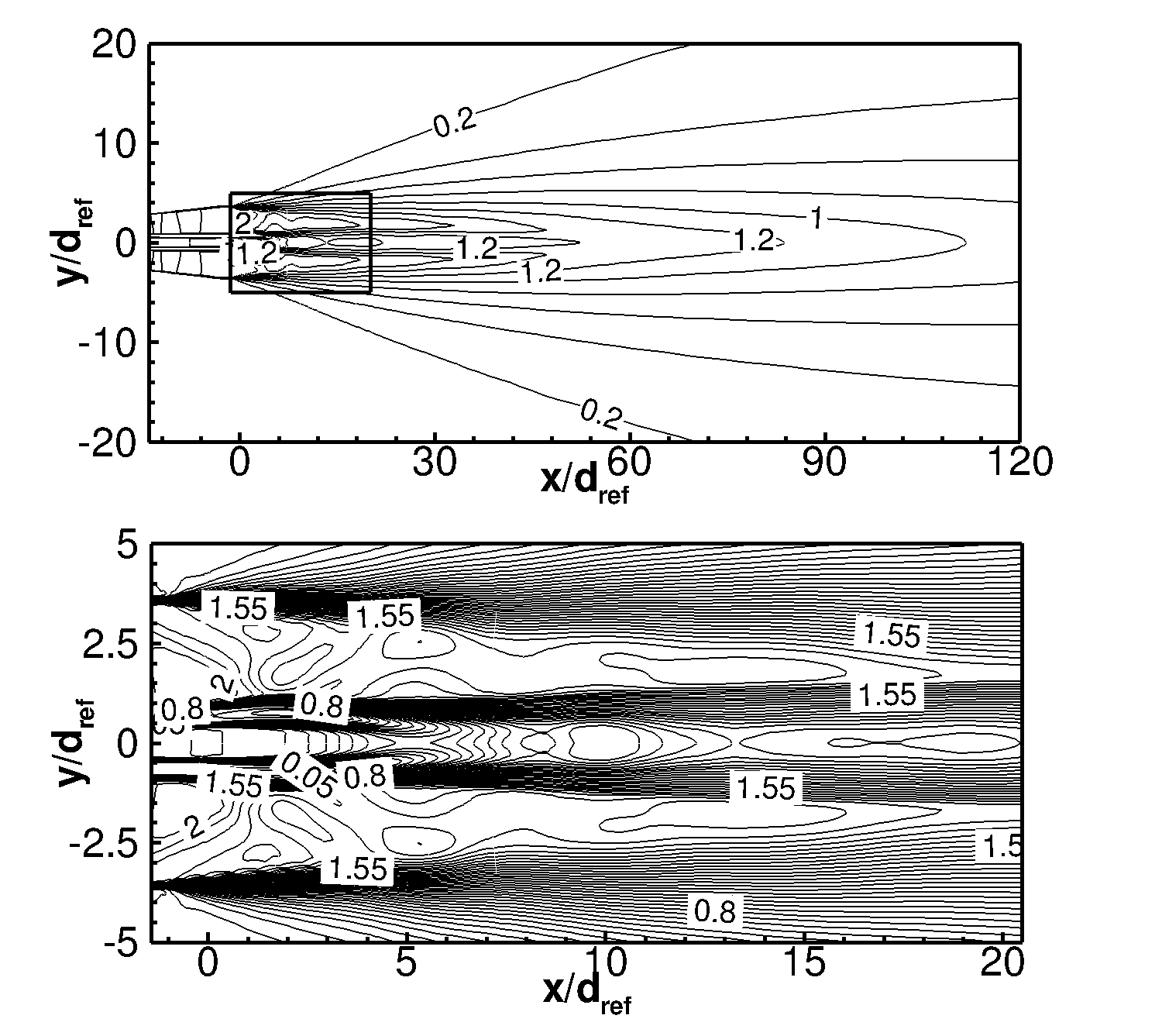}}
\subfigure[Model~B]{\includegraphics[scale=0.15]{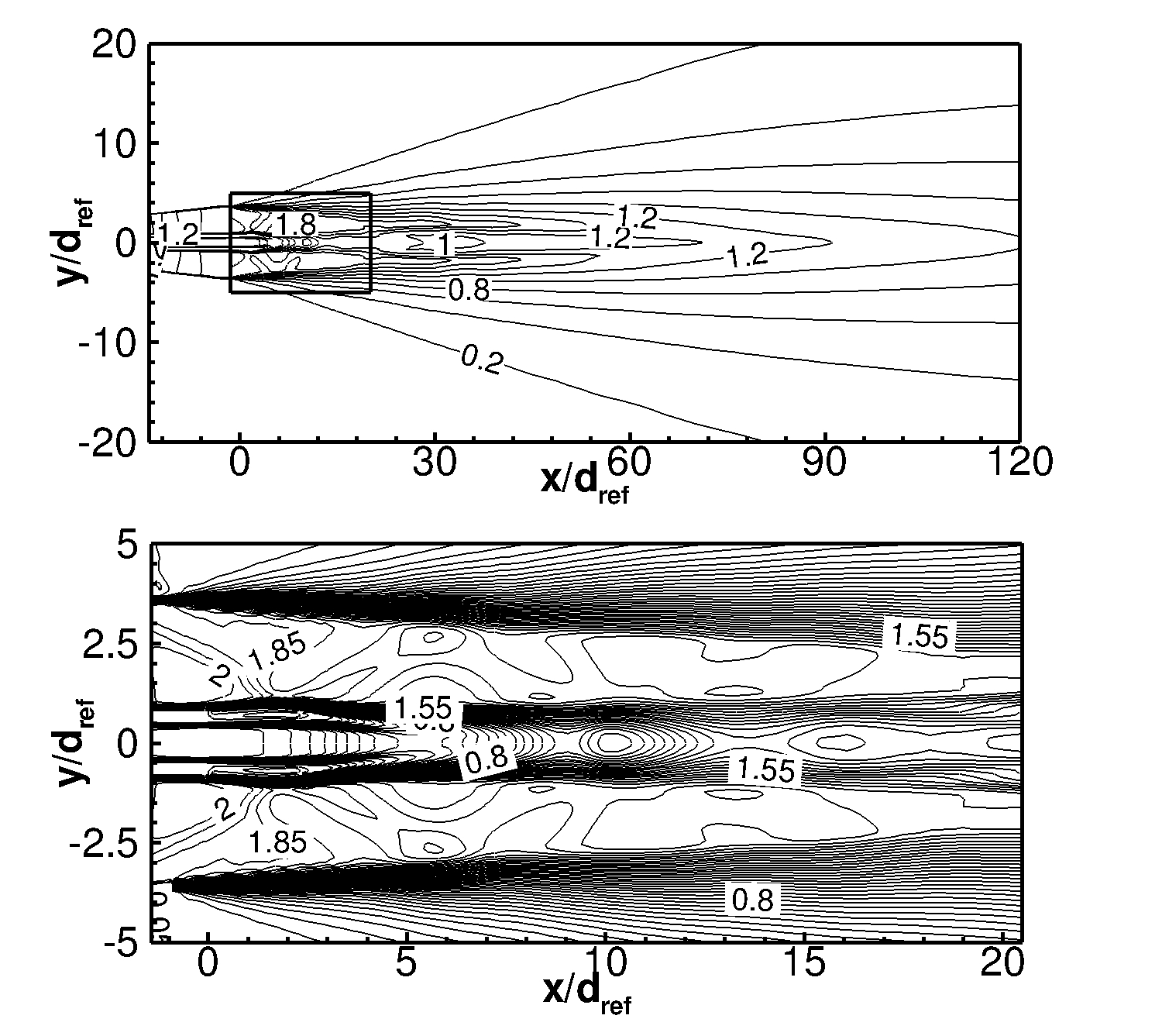}}
\caption{Mach number contours obtained with model~A (left) and model~B (right), with a close up of the near-burner region (bottom).}
\label{cheng_mach}
\end{center}
\end{figure}

\begin{figure}
\begin{center}
\includegraphics[scale=0.2]{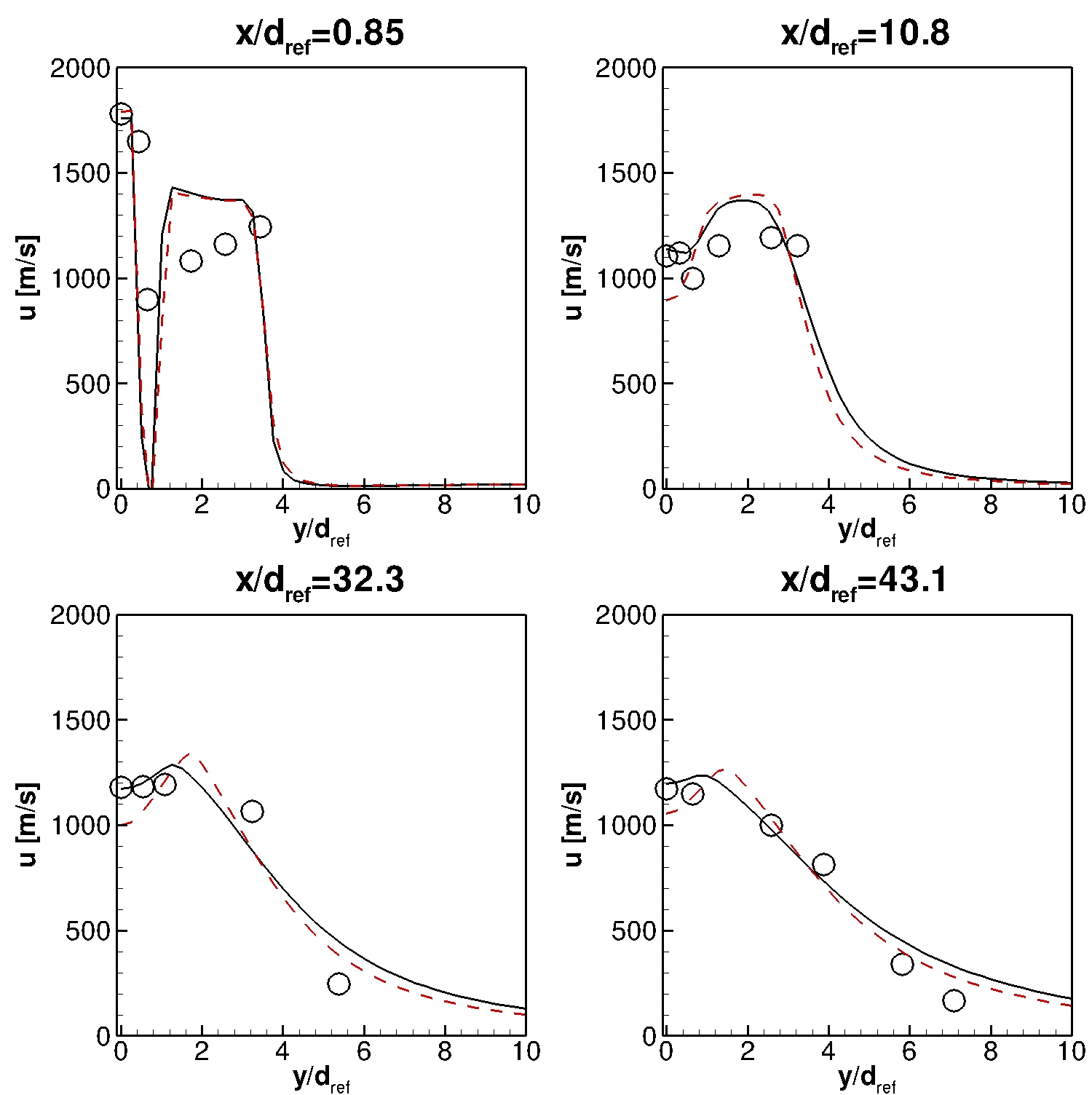}
\caption{Stream-wise velocity distributions at several radial sections: model~A, red dashed line; model~B, black solid line; symbols, experimental data.}
\label{cheng_vel}
\end{center}
\end{figure}

\begin{figure}
\begin{center}
\includegraphics[scale=0.2]{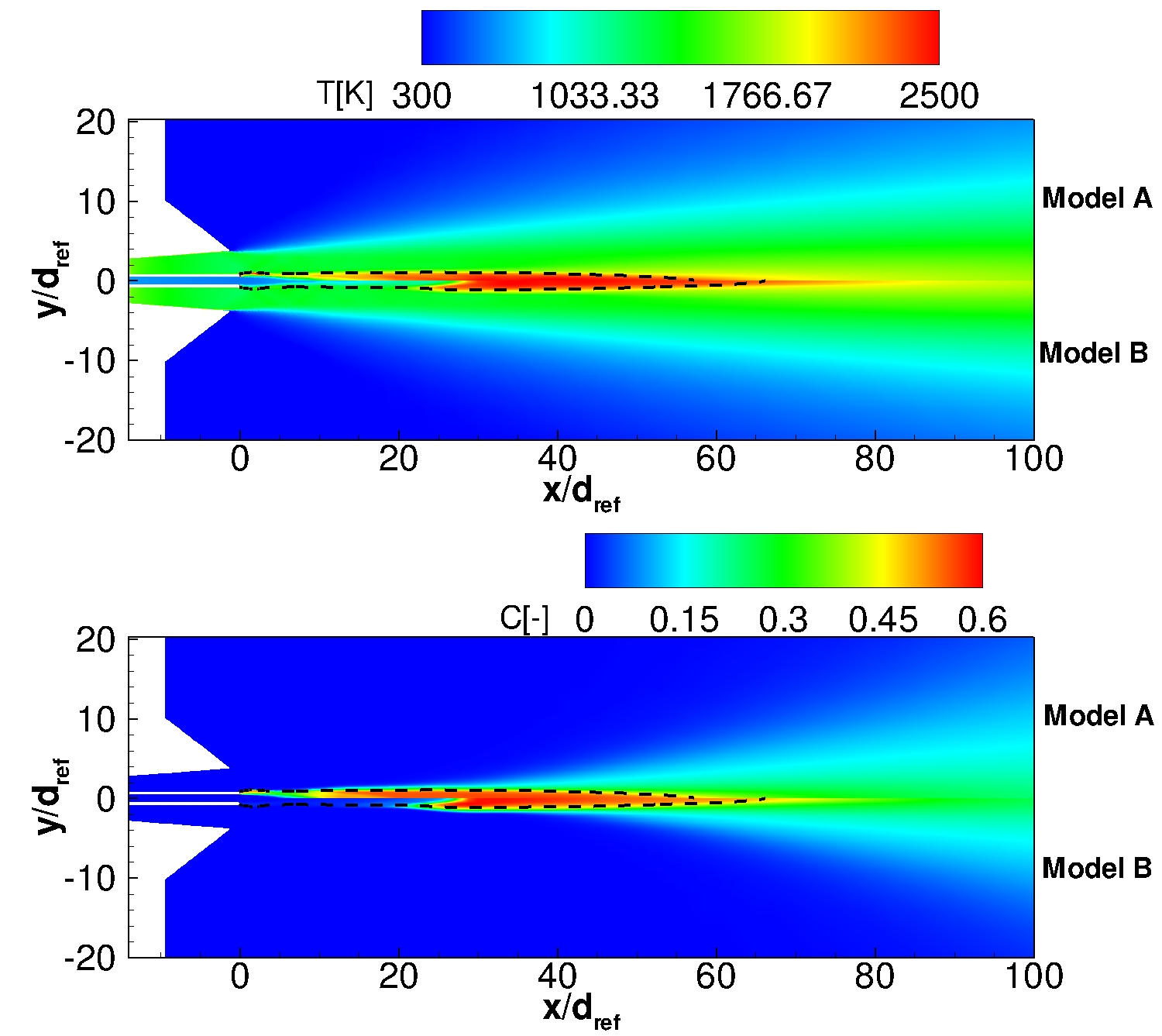}
\caption{Temperature (top) and progress variable (bottom) contours for the Cheng's combustion chamber with the $Z_{st}$ isoline superimposed (black dashed line).}
\label{cheng_temp_progvar}
\end{center}
\end{figure}

\begin{figure}
\begin{center}
\includegraphics[scale=0.2]{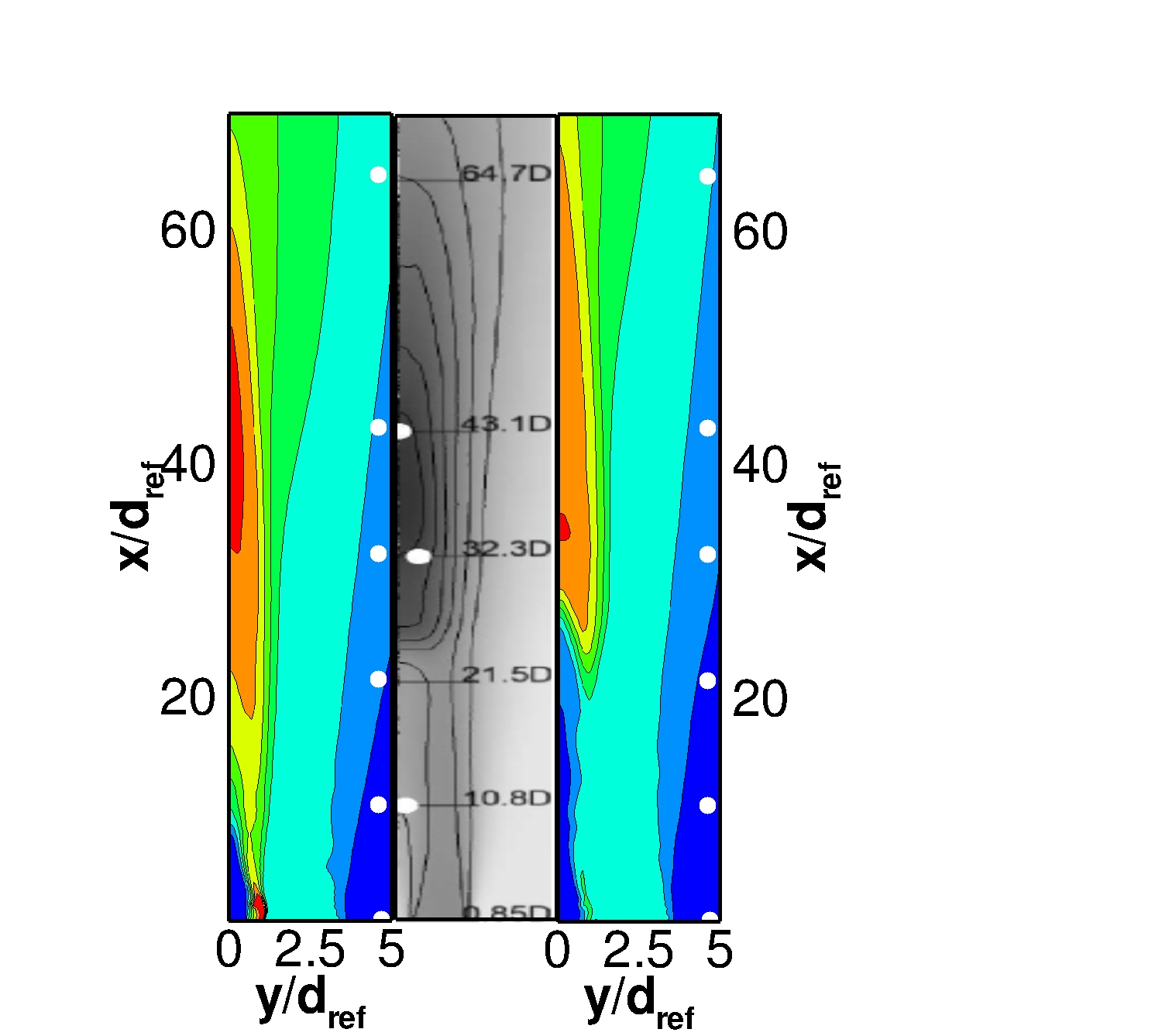}
\caption{Comparison between the temperature contours computed by model~A (left), Boivin et al.~\cite{boivin2012supersonic} (center), model~B~(right). The contour lines are plotted from $1000K$ to $2500K$ with step equal to $250K$.}
\label{cheng_cuenot}
\end{center}
\end{figure}

\begin{figure}
\begin{center}
\includegraphics[scale=0.32]{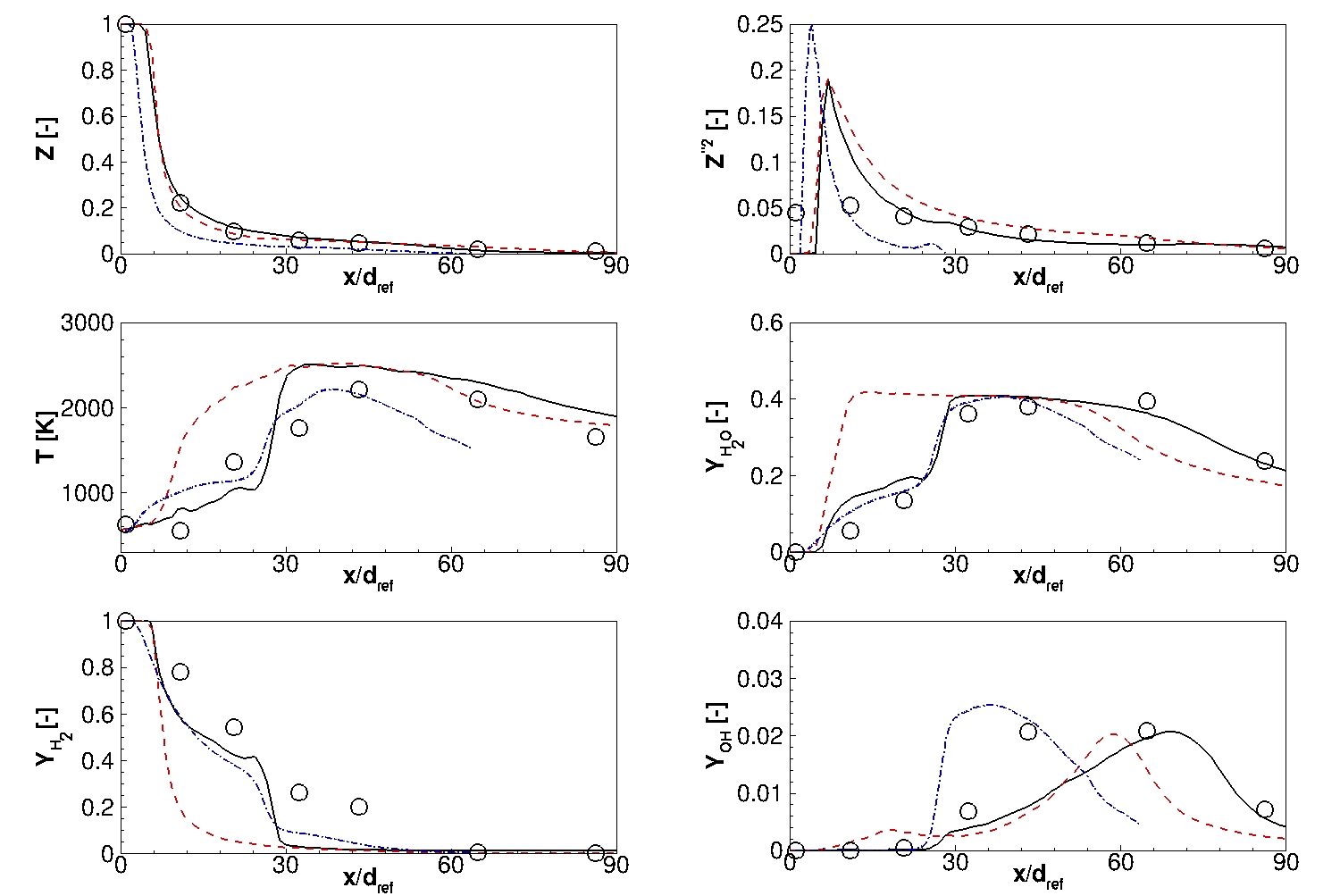}
\caption{Distributions of different thermo-chemical quantities} along the axis~$(y/d_{ref}=0)$. Model~A, red dashed line; model~B, solid black line; Boivin et al. dashed-dotted blue line~\cite{boivin2012supersonic}; symbols, experimental data~\cite{cheng1994}.
\label{cheng_axial}
\end{center}
\end{figure}

\begin{figure}
\begin{center}
\includegraphics[scale=0.3]{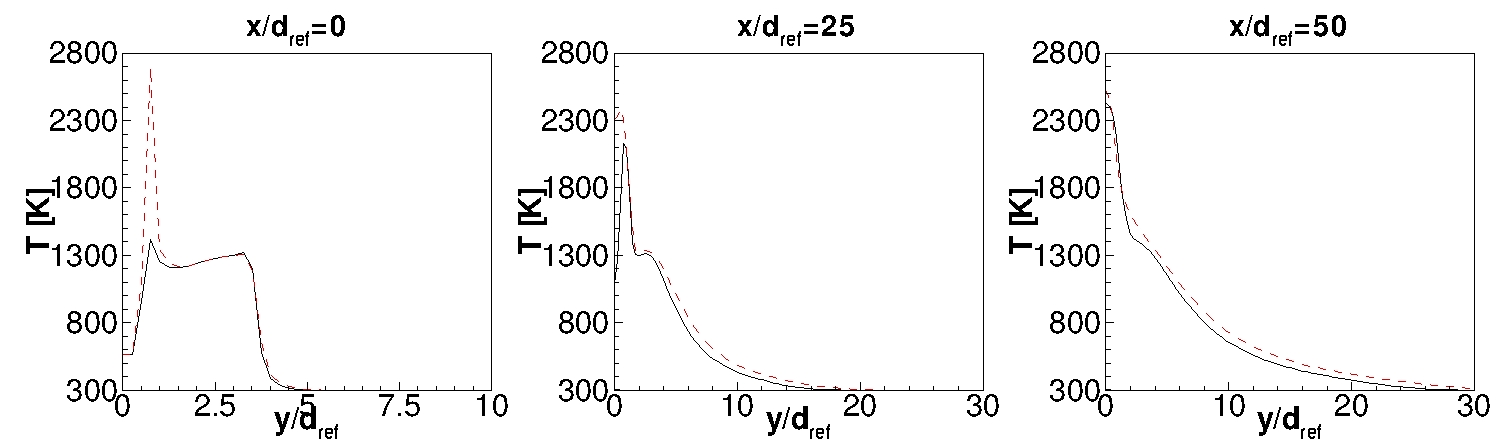}
\caption{Temperature distributions at $x/d_{ref}=0$, $x/d_{ref}=25$, and $x/d_{ref}=50$: model~A, red dashed line; model~B, solid black line.}
\label{cheng_slice}
\end{center}
\end{figure}

\begin{figure}
\begin{center}
\includegraphics[scale=0.3]{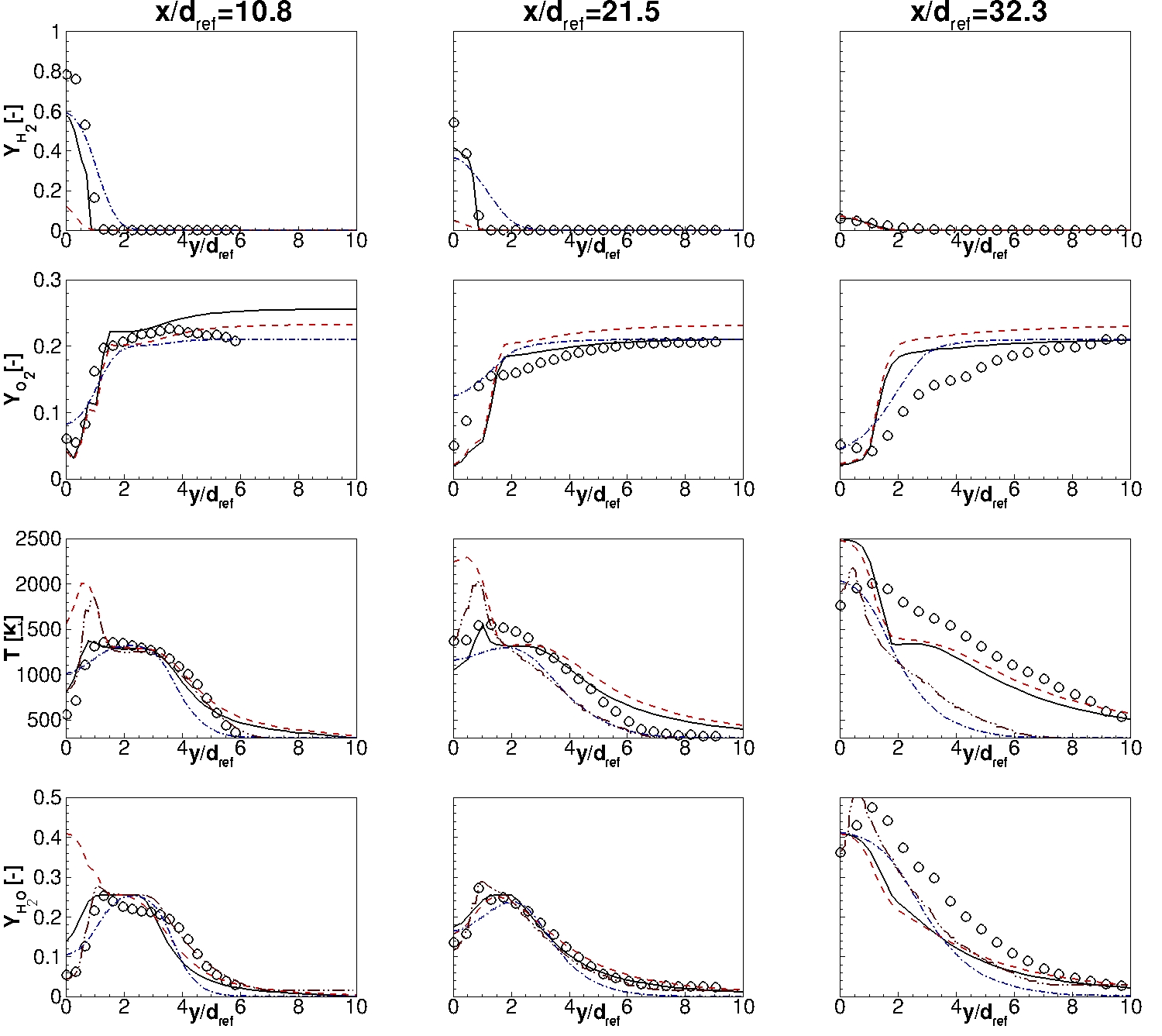}
\caption{Chemical species distributions at $x/d_{ref}=10.8$, $x/d_{ref}=21.5$, and $x/d_{ref}=32.3$: model~A, red dashed line; model~B, solid black line; Boivin et al. dashed-dotted blue line~\cite{boivin2012supersonic}; Thibault and Boivin dashed-dotted-dotted brown line~\cite{boivin2014CST}; symbols, experimental data~\cite{cheng1994}.}
\label{cheng_radial_CF}
\end{center}
\end{figure}

\begin{table}[h]
\centering
\caption[Parameter for the simulation of the Cheng's combustion chamber.]{Parameters for the simulation of the Cheng's combustion chamber~\cite{cheng1994}.}
\label{cheng_parameter}
\begin{tabular}{l|c c c c c c c c c c }
\hline
&$Ma$&$u$(m/s) & $T$ (K)& $p$ (kPa)& $Y_{H_{2}}$&$Y_{O_{2}}$&$Y_{H_{2}O}$&$Y_{N_{2}}$&$Tu_{in}$&$\nu_{T_{in}} (m^2/s)$\\
$H_{2}$ jet &1&1780&545&112&1&0&0&0&5.78$\%$&0.00023\\
\\
Air stream &2&1420&1250&107&0&0.201&0.255&0.544&10.24$\%$&0.00034\\
\hline
\end{tabular}
\end{table}
Figure~\ref{cheng_mach} provides the Mach number contours obtained using model~A and model~B. The close-up view of the near-burner region (bottom-left and -right panels) indicates that the inlet Mach number value is very close to the experimental data for both computations.
The distributions of the streamwise component of the velocity, $u$, at several abscissae along the chamber are shown in figure~\ref{cheng_vel}. 
A good agreement with the experimental data~\cite{cheng1994} is obtained for both models. 
Figure~\ref{cheng_temp_progvar}~(top frame) shows a qualitative comparison between the temperature contours evaluated with the two models.
It appears that the computed flame shapes are quite different. 
Model~A predicts a reaction zone attached to the burner, whereas model~B correctly predicts the flame detachment. 
In the bottom frame of figure~\ref{cheng_temp_progvar} one can find the two corresponding progress-variable contours.
The solution obtained using model~A provides non-zero values for the progress variable in the region close to the burner, indicating that the reaction is active. 
Figure~\ref{cheng_cuenot} presents the temperature field obtained using model~A~(left panel) and model~B~(right panel) in comparison with the results 
of an accurate LES by Boivin et al.~\cite{boivin2012supersonic}. 
One can observe that model~B predicts the lift-off height, evaluated as the position of maximum temperature gradient, at about $x=26\ d_{ref}$, whereas model~A predicts an height of about $9\ d_{ref}$. The results by Boivin et al.~\cite{boivin2012supersonic}, used as reference, predicts the stabilization at $25\ d_{ref}$. 
For both models, the jet shape and the aperture of the flame at $x/d_{ref}=50$, equal about to $10\ d_{ref}$, are in good agreement with the experimental data
(not shown).
For a quantitative analysis of the results, figures~\ref{cheng_axial}--\ref{cheng_radial_CF} provide the comparison among the results obtained using the two FPV models, the numerical results provided in~\cite{boivin2012supersonic,boivin2014CST}, and the experimental data~\cite{cheng1994}.
Figure~\ref{cheng_axial} shows the distribution of the main thermo-chemical quantities along the axis of the burner. The evaluation of the $\widetilde{Z}$ and $\widetilde{Z''^{2}}$ are very slightly improved by model~B; 
a reasonable improvement of the prediction of the distribution of $OH$ mass fraction is also obtained;
whereas, $H_2$ and $H_2O$ mass fractions are quite well predicted in comparison with model~A.
It appears that the flame core is not well reproduced using model~A, which indeed provides 
a too high reaction rate, so that the reactions may occur wherever the two flows ($H_{2}$--wet-air) are mixing. 
As a consequence, model~A predicts $Y_{H_2O}$ maximum at about $10\ x/d_{ref}$, as shown in
the middle right frame of figure~\ref{cheng_axial}.
On the other hand, model~B can reproduce the ignition spatial delay with a strong increase of the temperature at about $x/d_{ref}=30$ (figure~\ref{cheng_axial} middle-left frame). 

Figure~\ref{cheng_axial} also provides the numerical results obtained with an LES by Boivin et al.~\cite{boivin2012supersonic}, for comparison; they appear to be in good agreement with the results obtained by model~B, confirming the good level of accuracy achieved by such a model.
Analysing the temperature and water mass fraction experimental distributions shown in figure~\ref{cheng_axial}, one can notice that  
the temperature begins to decrease at $x/d_{ref}=50$ whereas the water mass fraction increases up to about $x/d_{ref}=65$. 
This is due to the entrainment of the surrounding unburnt-air into the flame, which induces an increase of the mass flow rate and
allows the reactions to occur even if the flame cannot heat up the surrounding flow.
In particular, the early increase of the temperature and of the water mass fraction, obtained by model A,
indicate that the heat is released very close to the burner. As a consequence, 
the entrainment of the surrounding unburnt-air into the flame is strongly overestimated using such a model with respect to model B.
The entrainment can be evaluated by computing the total mass flow rate through the inlet of the chamber, 
at $x/d_{ref}=0$, and at the sections $x/d_{ref}=25$ and $x/d_{ref}=50$. 
Due to the axial-symmetry, the effective surfaces intersected by the flow can be estimated as 
the circles with radius equal to the maximum extension of the temperature profiles, namely $(y/d_{ref})_0=5.540 $, $(y/d_{ref})_{25}=16.02 $, and $(y/d_{ref})_{50}=27.115$ (see figures~\ref{cheng_slice}). 
The mass fluxes computed using the solution of model~B are equal to $\dot m_{x/d_{ref}=0}=96.28~g/s$, $\dot m_{x/d_{ref}=25}=511.78~g/s$, and $\dot m_{x/d_{ref}=50}=1163.77~g/s$, respectively.
The different behaviour of model~A reflects in the evaluation of the mass fluxes which are: $\dot m_{x/d_{ref}=0}=98.07~g/s$, $\dot m_{x/d_{ref}=25}=681.83~g/s$, and $\dot m_{x/d_{ref}=50}=1202.08~g/s$. 

Figure~\ref{cheng_radial_CF} shows the distribution of some relevant thermo-chemical variables at several abscissae along the chamber axis, comparing the results obtained by the two FPV models, an LES with reduced chemistry evaluation~\cite{boivin2012supersonic}, a mixture-fraction based model~\cite{boivin2014CST}, and experimental results~\cite{cheng1994}.
The $H_2$ and $O_2$ mass fraction distributions agree fairly well with LES and experimental data. The improvement with respect to the results of model~A is more evident in the near axis region. Model~A predicts a thin reacting zone near the burner, so that in sections at $x/d_{ref}=0.85$~(see figure~\ref{cheng_slice}) and $x/d_{ref}=10.8$~(see figure~\ref{cheng_radial_CF}) the temperature distributions present a spike close to the axis. This spike characterizes also the water mass fraction distribution of the mixture-fraction-based model by Thibault and Boivin~\cite{boivin2014CST}. 
On the other hand, model~B is not affected by this problem and correctly predicts the flame core.
\subsection{The flame structure}
In this section the structure of the flame is studied with reference to the work of Moule et al.~\cite{moule2014} and of Boivin et al.~\cite{boivin2012supersonic} who have analyzed the considered supersonic burner.
Moule et al.~\cite{moule2014} identify three characteristic regions of the flame, namely: 1) the auto-ignition zone, $10\leq x/d_{ref}\leq 18$, where the mixture prepares to ignite; 2) the stabilization region, $18\leq x/d_{ref}\leq 26$, where the flame starts; 3) the combustion region, $30\leq x/d_{ref}\leq 40$.

In the first zone, the initial steps of the hydrogen oxidation take place and the reaction produces simple hydrogen radicals. 
In particular, for hydrogen combustion, the hydroperoxyl radical $HO_2$ can be considered as a good marker of autoignition. However,
since the $HO_2$ concentration also increases in fuel-rich reaction zones of ignited mixtures, 
Boivin et al.~\cite{boivin2012supersonic} proposed a more accurate criterion for detecting autoignition. Such a criterion requires
the simultaneous presence of high values of the normalized production rate of $HO_2$ and of the reactivity of the mixture, $\lambda$,
defined as the positive eigenvalue of the Jacobian of the chemical source term associated to the autoignition chain-branching 
reaction~\cite{boivin2012supersonic}.
For the present supersonic lift-off flame,
the results of the LESs discussed in references~\cite{moule2014} and \cite{boivin2012supersonic} confirm the effectiveness of
such a criterion in identifying the autoignition region as well as the following transition from the autoignition to the stabilization characterized by the $HO_2$ depletion.

\begin{figure}
\begin{center}
\subfigure[Model~A]{\includegraphics[scale=0.15]{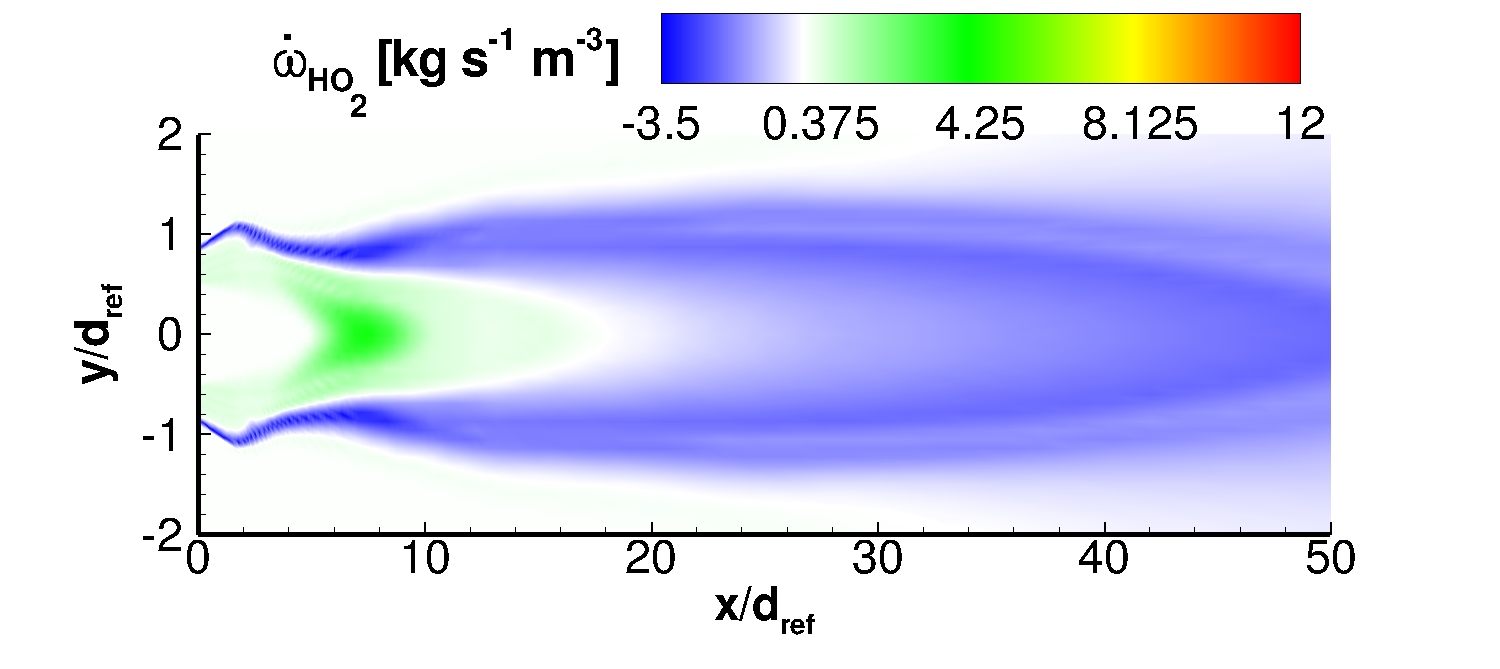}}
\subfigure[Model~B]{\includegraphics[scale=0.15]{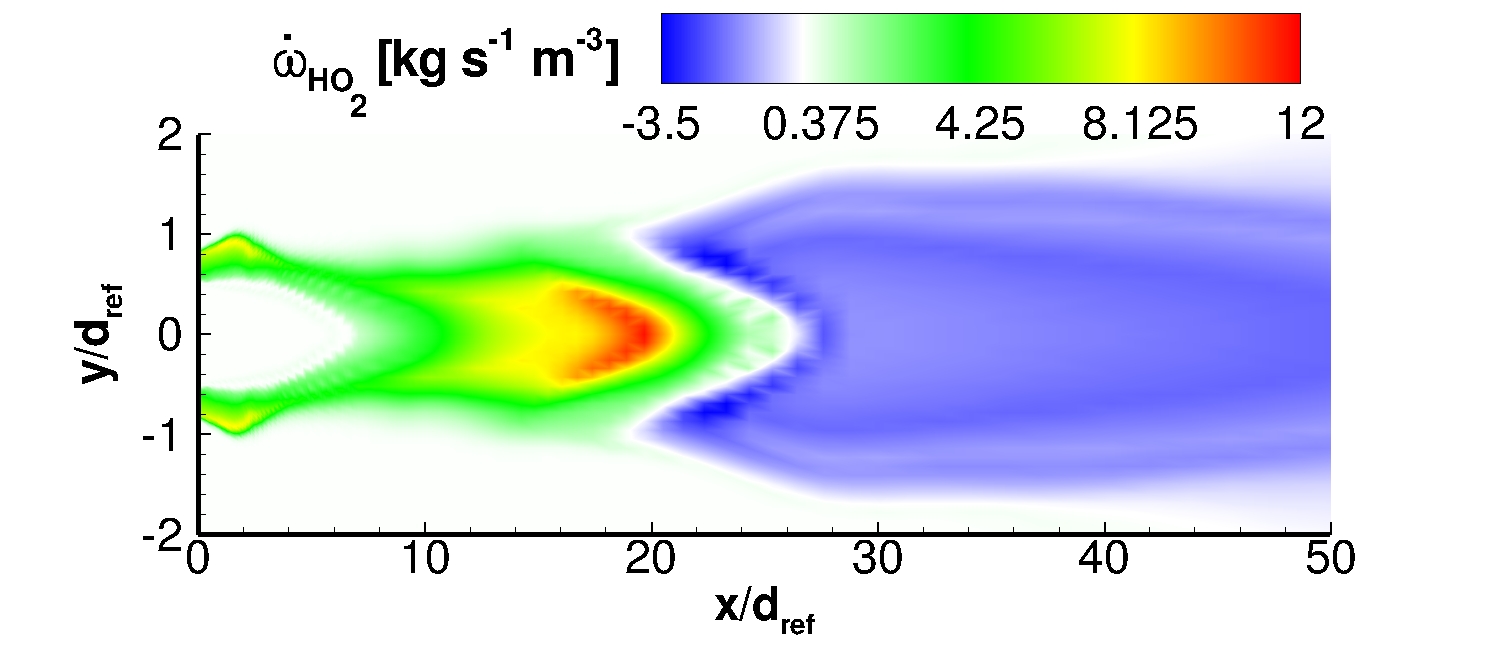}}
\caption{Model~A and Model~B solutions: contours of $\dot{\omega}_{HO_2}$.}
\label{cheng_HO2}
\end{center}
\end{figure}

\begin{figure}
\begin{center}
\subfigure[Model~A]{\includegraphics[scale=0.15]{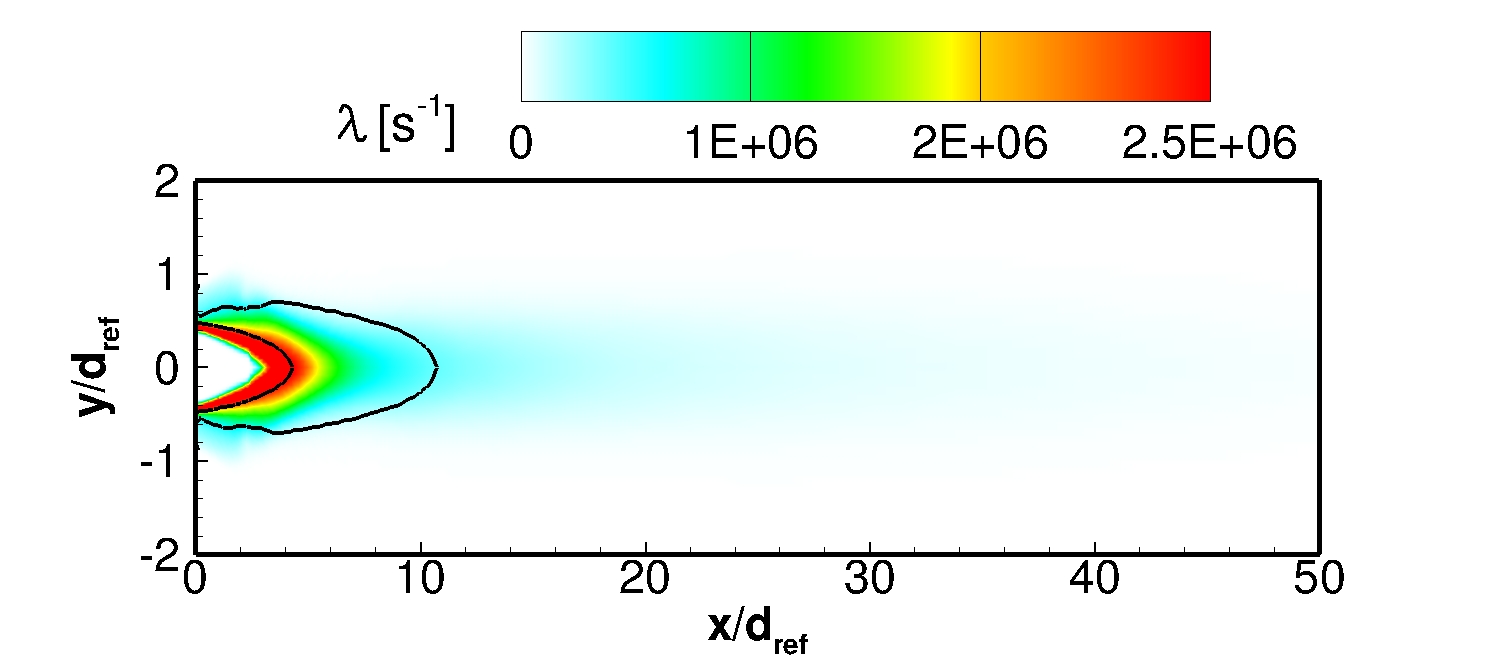}}
\subfigure[Model~B]{\includegraphics[scale=0.15]{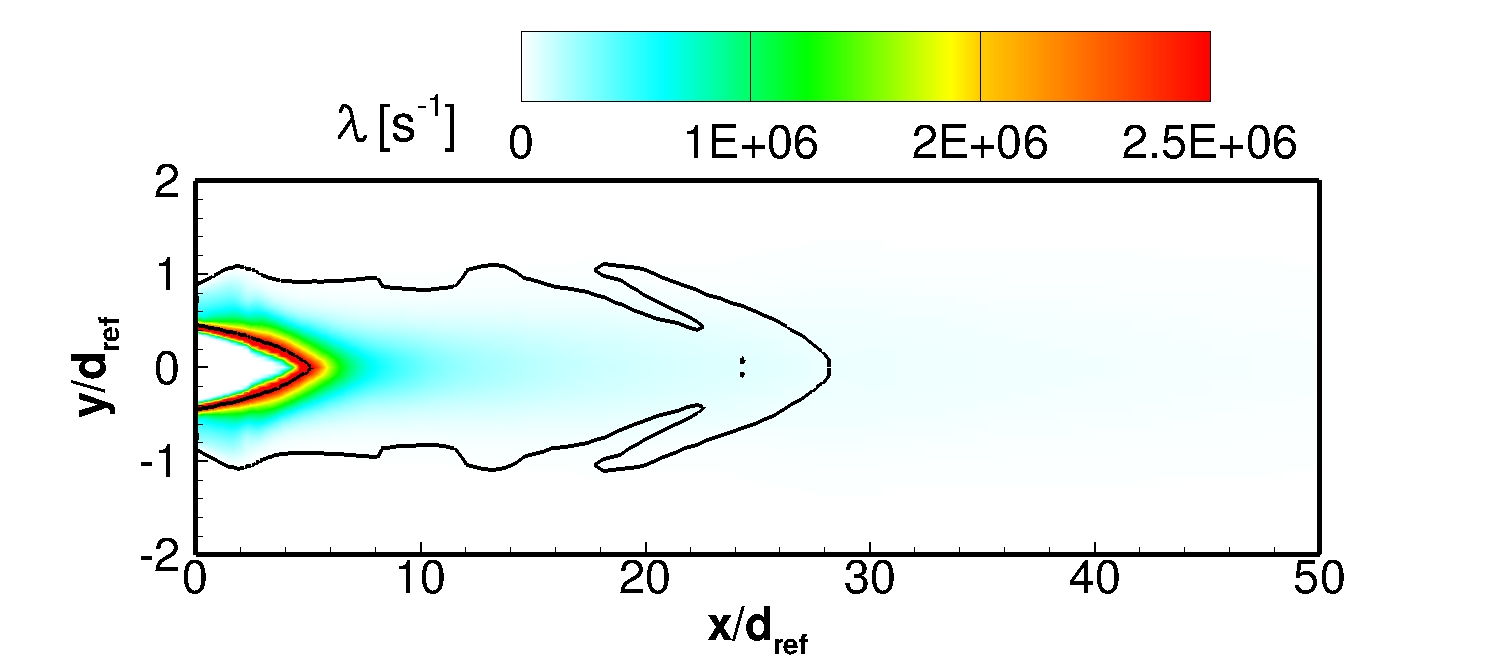}}
\caption{Model~A and Model~B solutions: reactivity contour plot with the $Y_{HO_2}$ isoline [$3\times 10^{-5}$] superimposed.}
\label{cheng_reactivity}
\end{center}
\end{figure}

Analysing the results of the present RANS simulations, we can show that model B can provide a flame structure corresponding
to the above scenario.
Figure~\ref{cheng_HO2} shows the contours of the $HO_2$ production rate, $\dot{\omega}_{HO2}$, computed using model A (left) and model B (right). This figure indicates that both models predict a substantial $HO_2$ production downstream of the jet exit, in agreement with the results of the LES by Moule et al.~\cite{moule2014} (see figure 17 of their paper).
Moreover, it appears that the production region is followed by depletion taking place in the stabilization region.
Figure~\ref{cheng_reactivity} shows the contour plot of $\lambda$ 
together with the $Y_{HO_2}=3\times 10^{-5}$ isoline, considered as the reference value for the high temperature hydrogen autoignition~\cite{moule2014}. 
It appears that the most reactive part of mixture is in the near-burner zone, where the mixing among the reactants is stronger.  Both models properly evaluate the reactivity, in good agreement with the results of the LES of \cite{moule2014} (see figure 19 therein).
Looking at the contours of $\dot{\omega}_{HO2}$ in figure~\ref{cheng_HO2} and at the contours of $\lambda$  in figure ~\ref{cheng_reactivity}, we can verify the conjecture of Boivin et al. concerning the autoignition region. 
In fact, we can see that model B predicts the autoignition at $7\leq x/d_{ref}\leq 23$ 
where sufficiently high values of the reactivity and of the $HO_2$ production are achieved. 
On the other hand, model A predicts a too fast reaction rate, providing a shorter autoignition region for $3\leq x/d_{ref}\leq 11$.
\begin{figure}
\begin{center}
\subfigure[Model~A]{\includegraphics[scale=0.15]{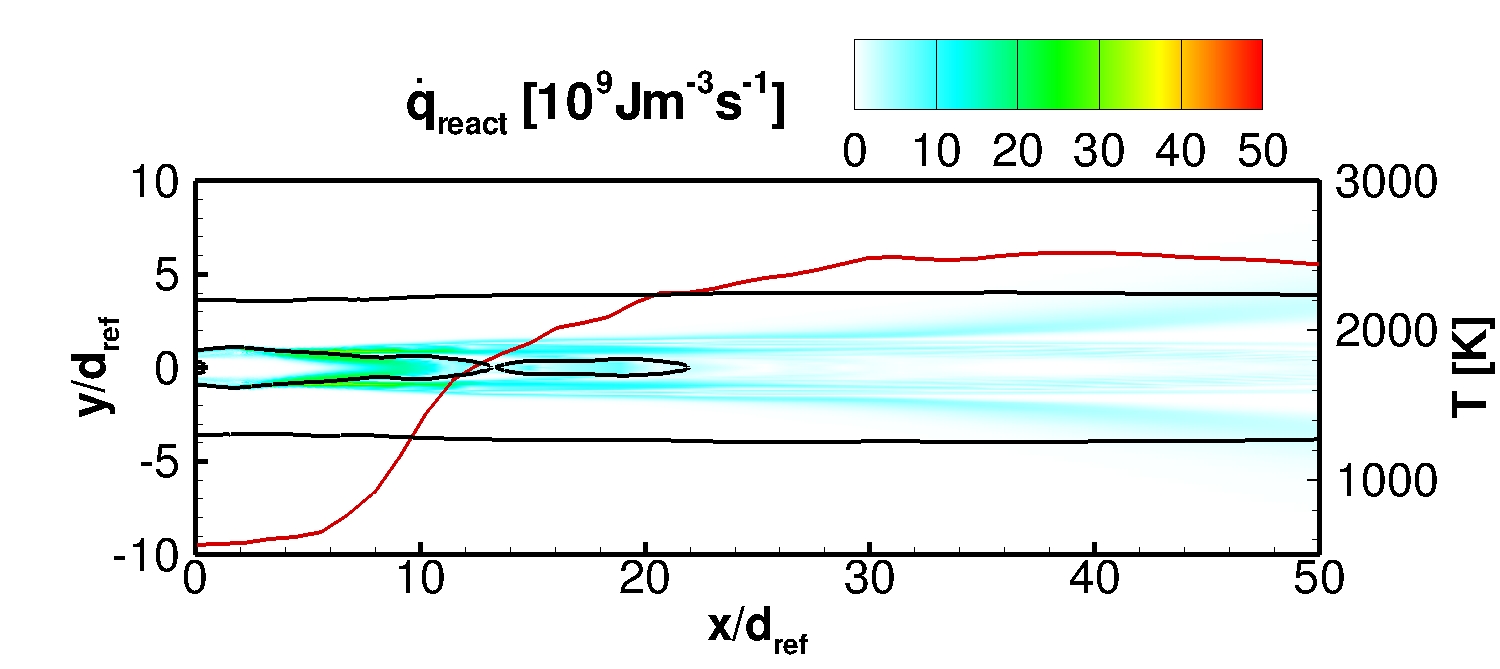}}
\subfigure[Model~B]{\includegraphics[scale=0.15]{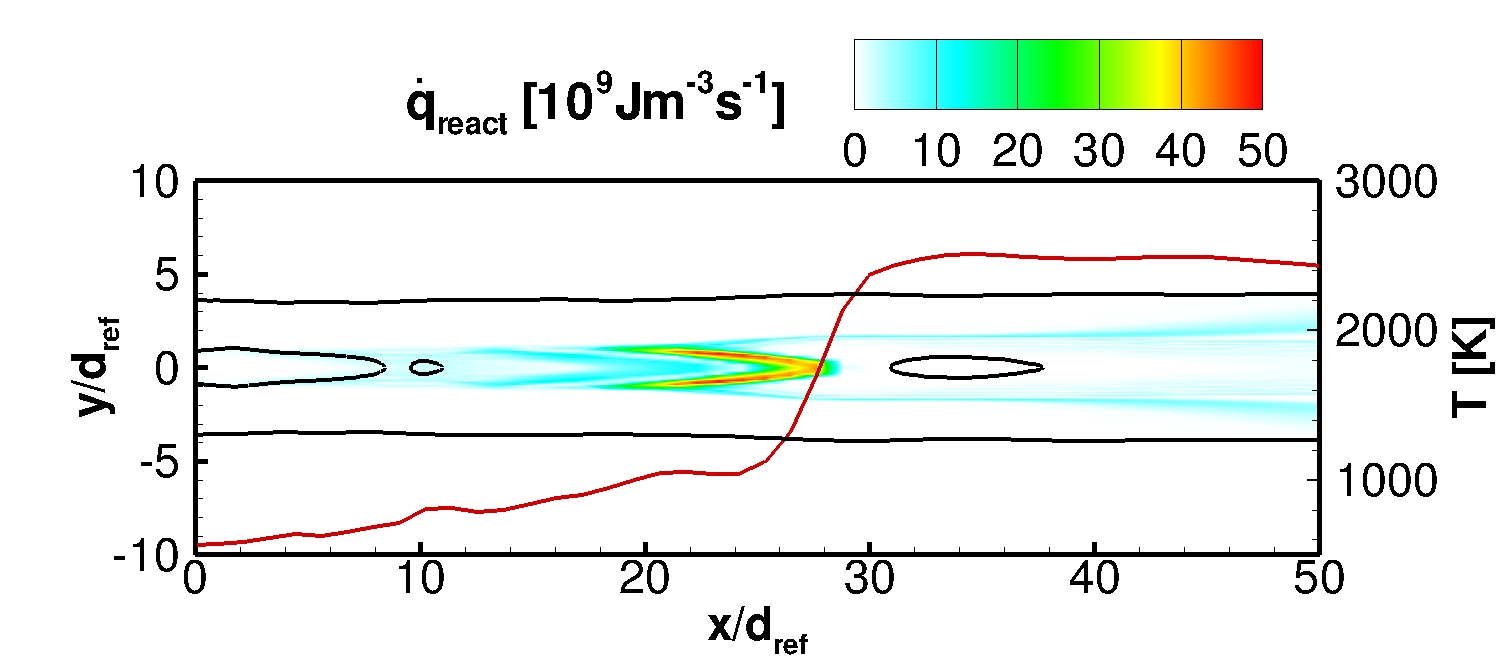}}
\caption{Model~A and Model~B solutions: heat reaction rate contour plot with the $Ma=1$ isoline (black solid line) and the axial temperature profile (red solid line) superimposed.}
\label{cheng_heat_release}
\end{center}
\end{figure}

In the second zone,  the flame finds its stable position and, 
as shown in figure~\ref{cheng_heat_release}, it is characterized by high values of the heat release rate.
Figure~\ref{cheng_heat_release} also shows the sonic isoline and the axial temperature profile superposed to the contours of the
heat release rate.  According to the LES results of~\cite{moule2014},
although the highest temperature occurs in the subsonic region, $32\leq x/d_{ref}\leq 38$ , 
high values of $\dot{q}_{react}$ are concentrated in the neighbourhood of the stabilization region of the flame, $18\leq x/d_{ref}\leq 26$. 
For the present RANS computations, model A predicts high values of the temperature and of the heat release too close to the burner,
whereas, model B provides results in very good agreement with the LES concerning both the heat release region and the
temperature profile along the axis of the burner. 

\begin{figure}
\begin{center}
\subfigure[Model~A]{\includegraphics[scale=0.15]{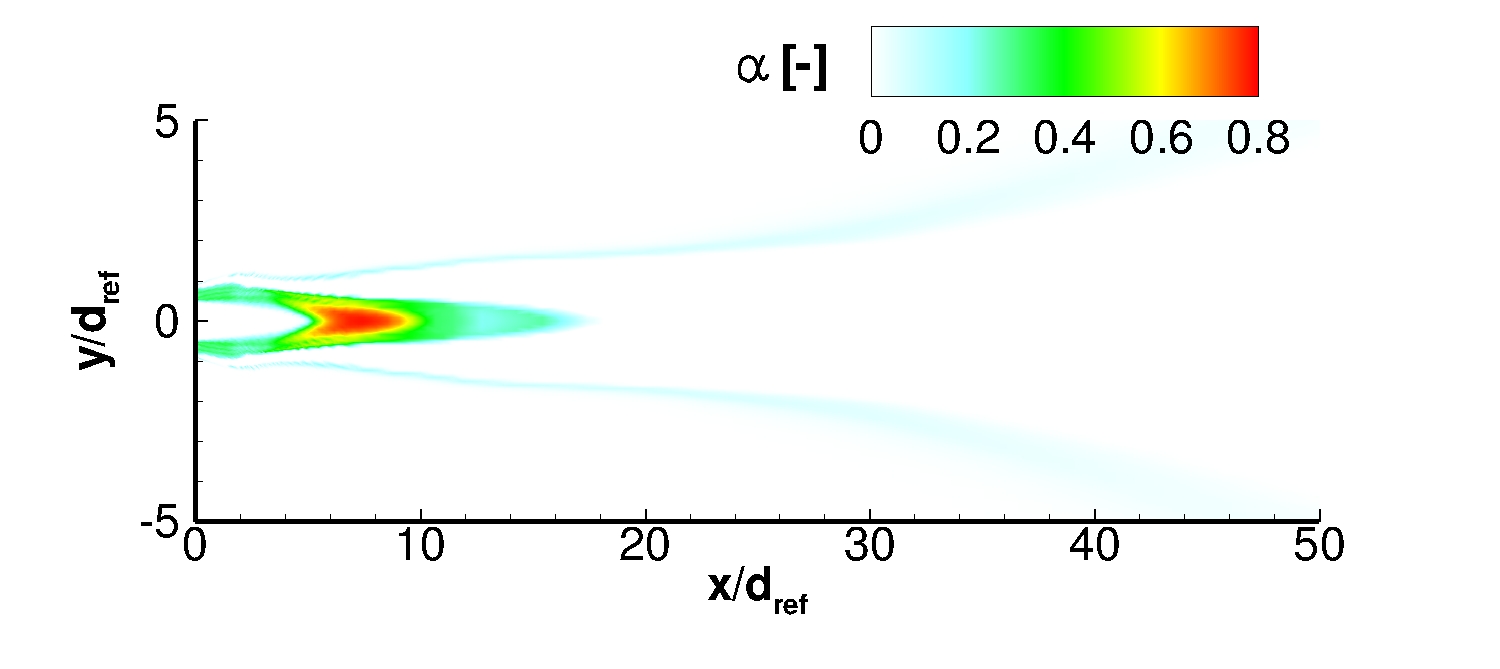}}
\subfigure[Model~B]{\includegraphics[scale=0.15]{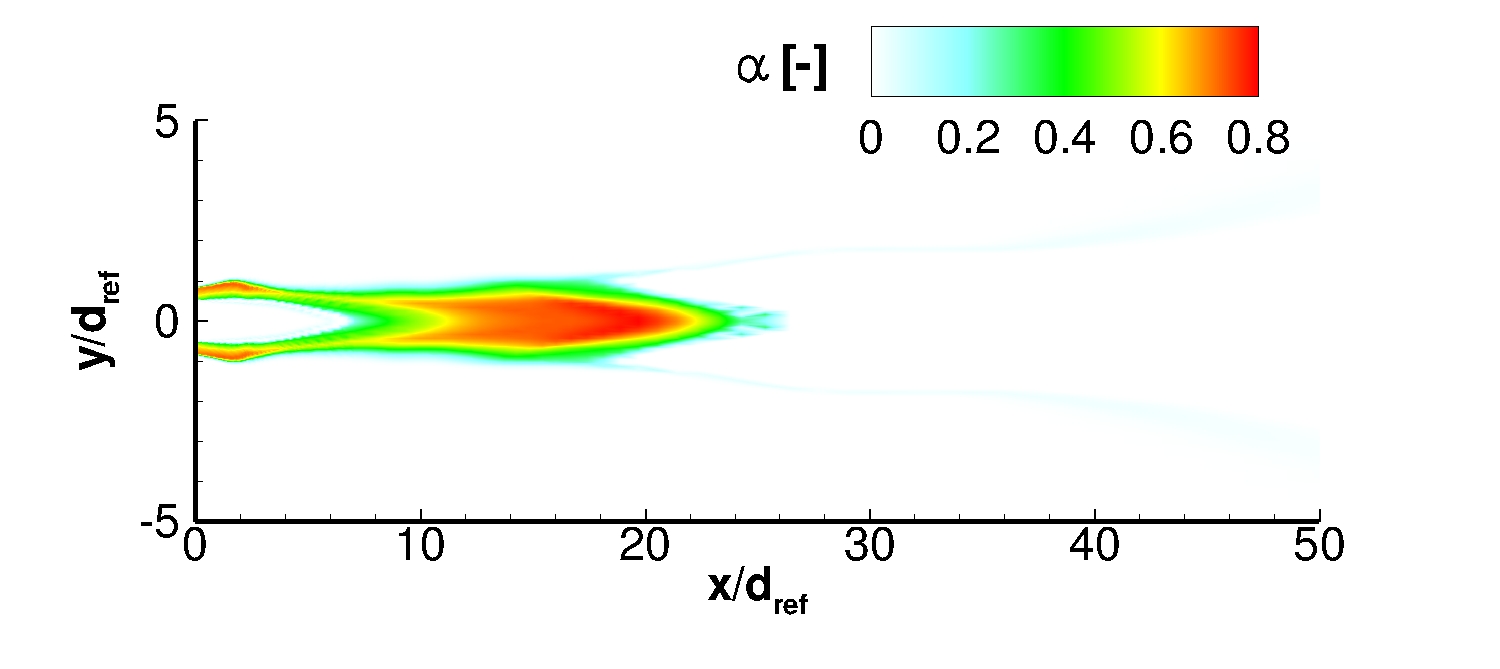}}
\caption{Model~A and Model~B solutions: contour plot of $\alpha$.}
\label{cheng_alpha}
\end{center}
\end{figure}

As proposed by Boivin et al.~\cite{boivin2012supersonic}, it is also possible to employ a unique parameter 
that identifies the occurrence of autoignition, 
\begin{equation}
\label{alpha}
{\alpha=\frac{\dot{\omega}_{HO_2}^+-\dot{\omega}_{HO_2}^-}{\dot{\omega}_{HO_2}^+}}\, ,
\end{equation} 
with $\dot{\omega}_{HO_2}^{+/-}$ the positive/negative part of $\dot{\omega}_{HO_2}$ representing the production and the destruction rates of $HO_2$. According to Boivin et al.~\cite{boivin2012supersonic},
the autoignition occurs when $\alpha$ decreases from its maximum reference value ($\alpha_{max}\approx 0.95$) to
its minimum reference value ($\alpha_{min}\approx 0.05$). 
Figure~\ref{cheng_alpha} shows the two $\alpha$ contour plots obtained by model~A and~B, respectively, 
confirming that model~B is capable of predicting the transition from the autoignition to the stabilization region at
about $27\ d_{ref}$, providing a good estimate of the lift-off height.

\begin{figure}
\begin{center}
\subfigure[Model~A]{\includegraphics[scale=0.15]{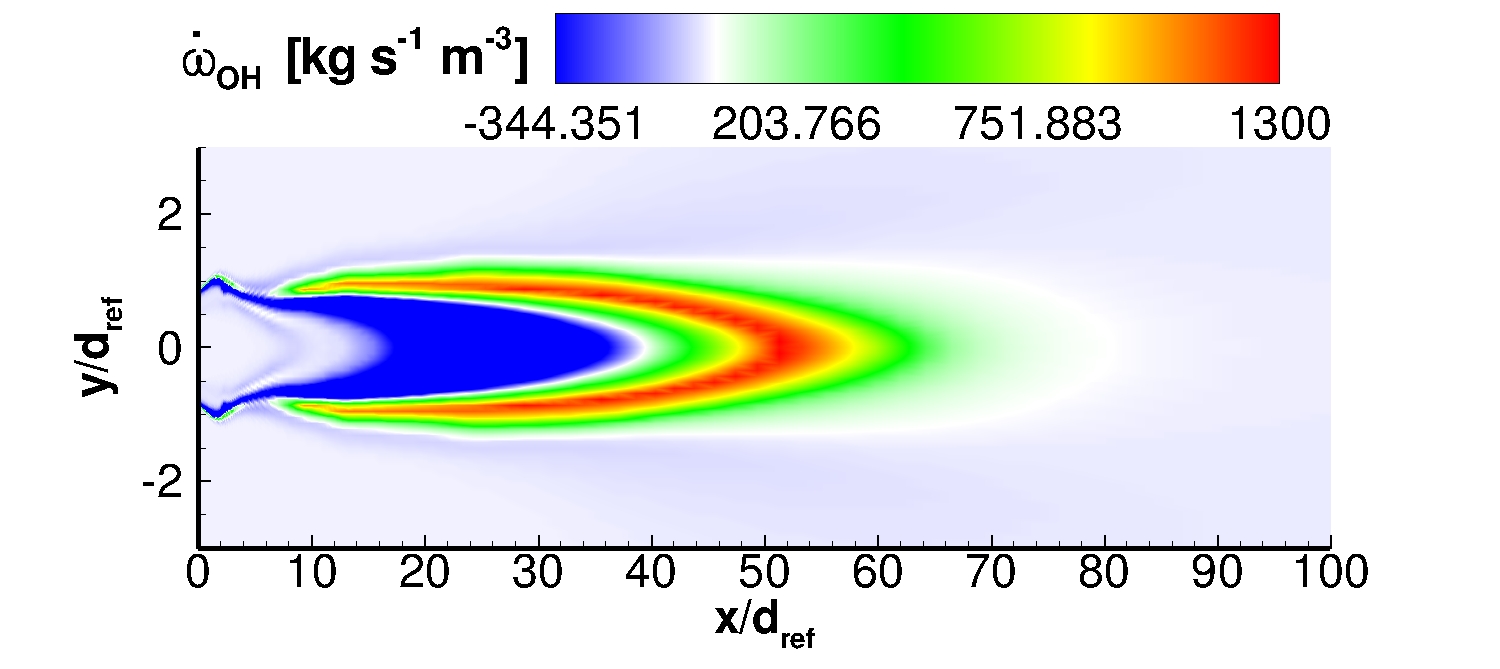}}
\subfigure[Model~B]{\includegraphics[scale=0.15]{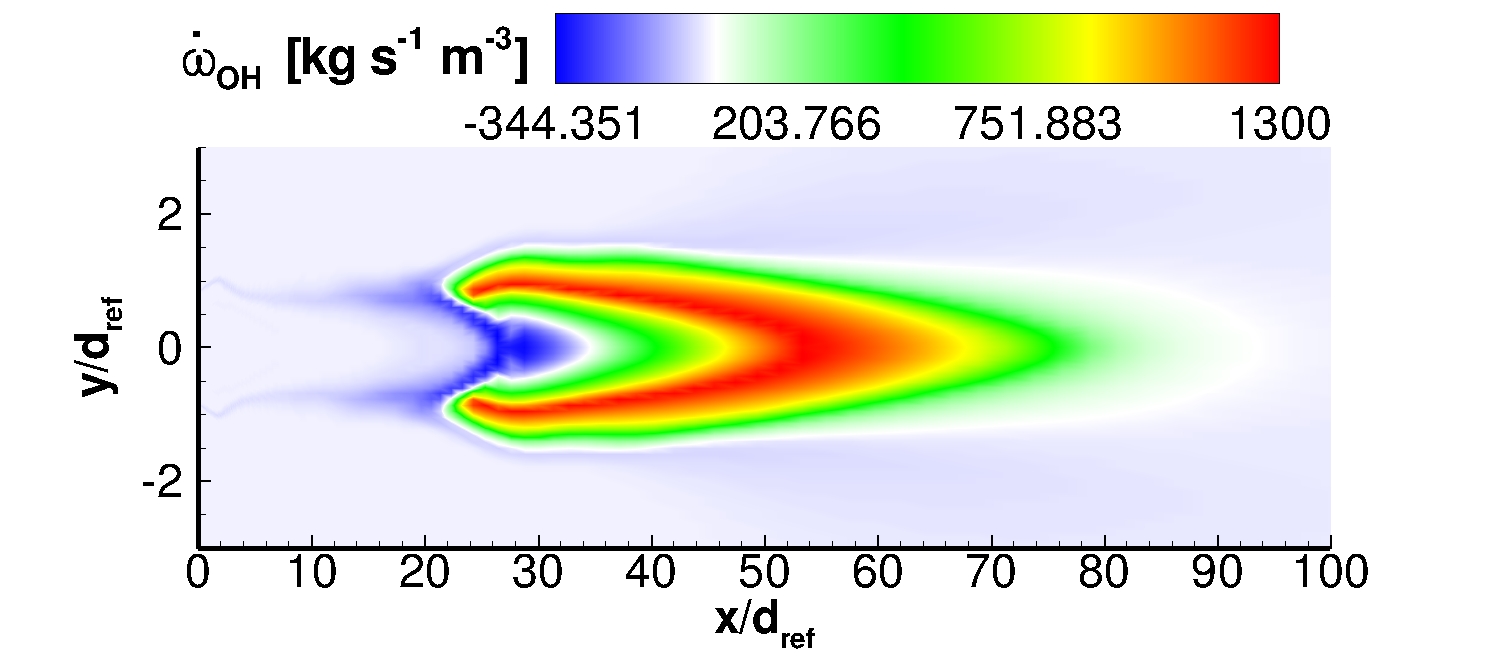}}
\caption{Model~A and Model~B solutions: contour plot of $\dot{\omega}_{OH}$.}
\label{third_zone}
\end{center}
\end{figure}

Finally, figure~\ref{third_zone} shows that in the third region ($30\leq x/d_{ref}\leq 40$) the combustion develops and is
characterized by high values of ${OH}$ production.


\subsection{$Z$ and $\Lambda$ distributions}

\begin{figure}
\begin{center}
$\widetilde{Z}=0.34,\, \widetilde{Z''^2}=0.024,\, \widetilde{\Lambda}=0.008,\, \widetilde{\Lambda''^2}=0.0$\\
\includegraphics[scale=0.125]{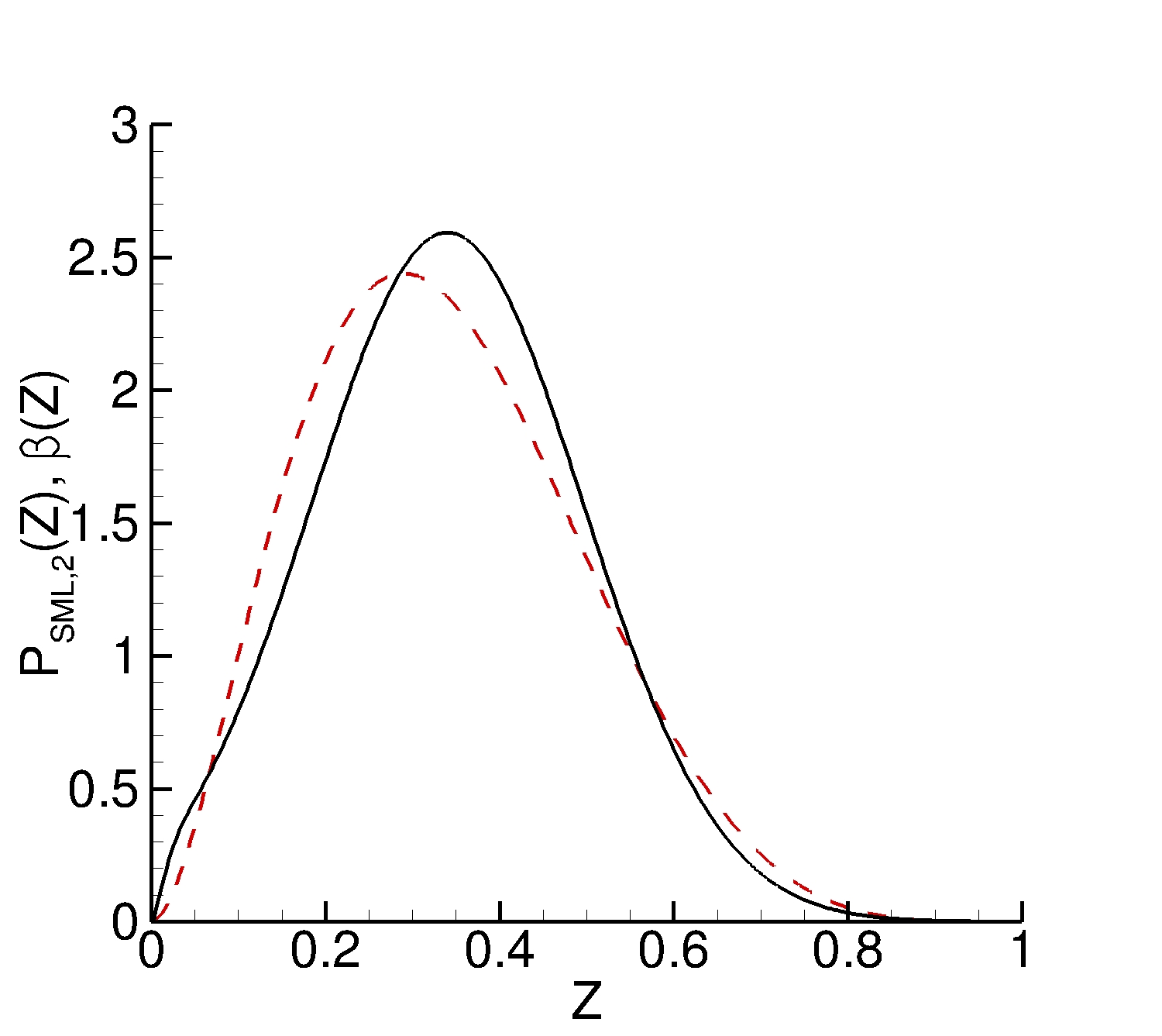}
\caption{PDFs of $Z$ at the point $(x/d_{ref}, y/d_{ref})=(8,0)$ using the values obtained by model~B:
$P_{SML,2}(Z)$ (black solid line) and $\beta(Z)$ (red dashed line). }
\label{cheng_pdfx8}
\end{center}
\end{figure}
\begin{figure}
\begin{center}
$\widetilde{Z}=0.065,\, \widetilde{Z''^2}=0.0012,\, \widetilde{\Lambda}=0.265,\, \widetilde{\Lambda''^2}=0.058$\\
\includegraphics[scale=0.35]{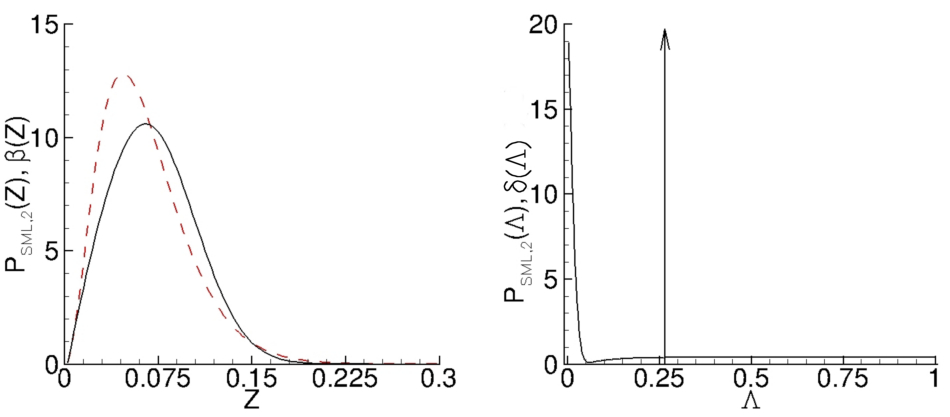}
\caption{PDFs of $Z$ and $\Lambda$ at the point $(x/d_{ref}, y/d_{ref})=(27,0)$ using the values obtained by model~B. Left panel:
$P_{SML,2}(Z)$ (black solid line) and $\beta(Z)$ (red dashed line). Right panel: $P_{SML,2}(\Lambda)$ (black solid line) and $\delta(\Lambda)$.}
\label{cheng_pdfx27}
\end{center}
\end{figure}
\begin{figure}
\begin{center}
$\widetilde{Z}=0.21,\, \widetilde{Z''^2}=0.005,\, \widetilde{\Lambda}=0.0215,\, \widetilde{\Lambda''^2}=0.021$\\
\includegraphics[scale=0.35]{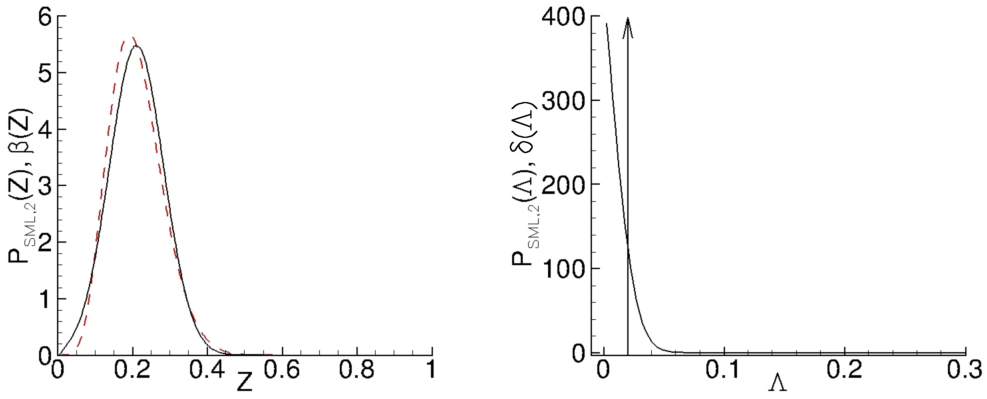}
\caption{PDFs of $Z$ and $\Lambda$ at the point $(x/d_{ref}, y/d_{ref})=(0.85,0.8)$ using the values obtained by model~B. Left panel:
$P_{SML,2}(Z)$ (black solid line) and $\beta(Z)$ (red dashed line). Right panel: $P_{SML,2}(\Lambda)$ (black solid line) and $\delta(\Lambda)$.}
\label{cheng_pdfx085_y08}
\end{center}
\end{figure}
\begin{table}[h]
\centering
\caption{Mean and variance of $Z$ and $\Lambda$ at three selected points.}
\begin{tabular}{ll|llll|}
                                                 &  & $\widetilde{Z}$  &$\widetilde{Z''^2}$  &$\widetilde{\Lambda}$  & $\widetilde{\Lambda''^2}$ \\
\midrule                                                 
\multirow{2}{*}{$(x/d_{ref}, y/d_{ref})=(8,0)$}  & Model~A & 0.37 & 0.030 & 0.22 & -- \\
                                                 & Model~B & 0.34 & 0.024 & 0.008 & 0 \\
\midrule
\multirow{2}{*}{$(x/d_{ref}, y/d_{ref})=(27,0)$} & Model~A & 0.085 & 0.0022 & 0.56 & -- \\
                                                 & Model~B & 0.065 & 0.0012 & 0.265 & 0.058 \\
\midrule
\multirow{2}{*}{$(x/d_{ref}, y/d_{ref})=(0.85,0.8)$} & Model~A & 0.22 & 0.007 & 0.49 & -- \\
                                                 & Model~B & 0.21
                                                  & 0.005 & 0.0215 & 0.021 \\
\midrule
\end{tabular}
\label{cheng_table}
\end{table}
In this section, we focus on the analysis of those flow regions where large differences between the predictions
of  model A and B are observed.
In particular, we compare the  $\widetilde{P}_{SML,2}$ with
the $\beta-$distribution for Z and with the $\delta-$function for $\Lambda$. All the distributions are 
evaluated using the values of the mean and variance computed by model~B at
three points: two of them are on the axis ($y/d_{ref}=0$)
and correspond to the lift-off heights evaluated with the two models, $x/d_{ref}=8$ and $x/d_{ref}=27$, respectively; 
the third point is close to the jet exit ($x/d_{ref}=0.85$, $y/d_{ref}=0.8$), where
a spurious temperature peak is predicted by model A (see figures~\ref{cheng_slice}). 

The first point ($x/d_{ref}=8$) is in the stabilization region for the solution obtained by model~A 
and falls in the mixing non-burning region for the solution obtained by model~B. 
This is clearly indicated by the mean and variance values of $Z$ and $\Lambda$ provided in table~\ref{cheng_table}: the
mean value of the progress parameter is 0.22 for model A, whereas it is very close to zero for model B. The values of the
mixture fraction are very close to each other.
Figure~\ref{cheng_pdfx8} shows the corresponding mixture-fraction distributions evaluated by the $\beta$-function and by
the $P_{SML,2}$ PDF.
The two distribution of $\Lambda$ (not shown) are both Dirac distributions since, due to the zero variance, the
$\beta$ function of model B collapses on a $\delta$-function centred at $\widetilde{\Lambda}=0.008$. 

At the second point on the axis of the burner ($x/d_{ref}=27$), the reaction is active for both models, as shown by
the progress parameter values in table~\ref{cheng_table}, which indicates that the combustion rate is higher for model A.
Model B also provides lower values for the mean and variance of the mixture fraction.
The PDF of $Z$ and $\Lambda$ are shown in figure~\ref{cheng_pdfx27}.  It appears that the $P_{SML,2}$ function is close to
the $\beta$-distribution due to the low variance.
The main differences between the two models are clearly observed
in the right panel of figure~\ref{cheng_pdfx27} where the $P_{SML,2}(\Lambda)$ and the $\delta(\Lambda)$ are shown in order to 
put in evidence the dramatic simplification made when assuming the Dirac distribution for the progress parameter.

Finally, considering the third point close to the burner ($x/d_{ref}=0.85$, $y/d_{ref}=0.8$), table~\ref{cheng_table} shows
that a large difference in the progress-variable mean value exists between the two models. In fact, model A provides
$\widetilde{\Lambda}=0.49$, indicating that combustion is active, whereas the mean value of $\Lambda$ for model B is close to zero.
Figure~\ref{cheng_pdfx085_y08} shows the corresponding PDF of $Z$ and $\Lambda$. Again, the $P_{SML,2}(Z)$ is very close
to the $\beta$-distribution (left panel), whereas the $P_{SML,2}(\Lambda)$ maintains a smoother behaviour with respect to
the $\delta$-function adopted for model A (right panel). The latter difference can be considered a reasonable motivation for the 
smoother behaviour of the temperature predicted close to the burner by model B (see figures~\ref{cheng_slice} and \ref{cheng_radial_CF}).

\section{Conclusions}

This paper presents a statistical more likely distribution (SMLD) approach for the evaluation of the presumed probability density function (PDF) in flamelet progress variable (FPV) models for non-premixed combustion. The FPV model is employed in conjunction with the Reynolds averaged Navier--Stokes (RANS) equations.
The proposed SMLD model is built evaluating the most probable joint distribution of the mixture fraction and of the progress variable and its adequateness and feasibility are discussed in comparison with the standard FPV model.
The SMLD model relies on a more robust theoretical basis with a
substantially unchanged computational cost.
Although the classical formulation of the FPV approach is based on the low-Mach-number assumption,
we solve the total energy conservation equation to improve the
computation of compressible reacting flows.
The performance of the two FPV models
are discussed by analysing the results of the simulation of a supersonic  $H_{2}$--Air combustion 
studied at the NASA Langley Research Center. 
This analysis shows that the FPV-SMLD model provides an effective improvement over the standard approach and is able to properly describe the flame structure in good agreement with the results obtained by highly resolved LES with detailed chemistry.  
In fact, the numerical results correctly predict the presence of the three characteristic regions of supersonic flames: 
the autoignition, the stabilization and the combustion regions. 
Moreover, the FPV-SMLD model can correctly evaluate the lift-off of the flame whereas 
the standard model is not able to predict
the combustion kinetic with sufficient accuracy, providing a sudden ignition of the mixture close to the burner inlet.
A detailed analysis of the PDF distributions at several points of the computational domain is provided in order to quantify and explain
the differences between the two models.
This work has shown that indeed evaluating the PDF using the SMLD approach allows one to improve the accuracy of
the simulation of a reacting supersonic flow in the framework of RANS equations. However, several aspects should be considered
to enhance this analysis, such as: the effects of compressibility deserve to be analyzed in more details employing, for instance, a compressible 
FPV model;
the effects of the hypothesis that the mixture fraction and the progress variable are not statistically independent have to be studied; 
the influence of the proposed FPV-SMLD model in the LES framework should be investigated. However, all this topics go beyond the aim
of the present paper and will be the subject of our future work.

\section*{Acknowledgements}
This research has been supported by grant n. $PON03PE\_00067\_6$ APULIA SPACE.
The authors are very grateful to M. D. de Tullio for his support in the software development and to T.S. Cheng for providing the experimental data used as reference. Then, the authors wish to thanks P. Boivin for providing numerical LES results used as reference and for the interesting and useful discussion. 
\bibliographystyle{elsarticle-num.bst}
\bibliography{biblio.bib}

\end{document}